\documentclass[a4paper,11pt]{article}
\usepackage{authblk}

\usepackage[T1]{fontenc}

\usepackage[utf8]{inputenc}
\usepackage[margin=30mm]{geometry}
\usepackage{changepage}

\usepackage{textcomp}
\usepackage{float}
\usepackage{amsmath, amssymb}
\usepackage{braket}
\usepackage[english]{babel}
\usepackage[separate-uncertainty=true]{siunitx}
\usepackage{graphicx}
\usepackage[labelfont=bf]{caption}
\usepackage{subcaption}
\usepackage{mathtools}
\usepackage{physics}
\usepackage{amssymb}
\usepackage{amsmath}
\usepackage{mathtools}
\usepackage{xfrac}
\usepackage{array}
\usepackage{relsize}
\usepackage{tabu}
\usepackage{bbold}

\usepackage[font=small]{caption}
\usepackage{amsfonts}
\usepackage{booktabs}
\usepackage{siunitx}
\usepackage{appendix}
\usepackage{mathrsfs}
\usepackage{amsthm}
\usepackage{indentfirst}
\usepackage[nottoc,numbib]{tocbibind}
\usepackage[autostyle,italian=guillemets]{csquotes}
\usepackage[style=numeric,backend=biber,sorting=none]{biblatex}

\renewbibmacro{in:}{}

\title{Stress-energy tensor correlations across regular black holes horizons}
\author{Matteo Fontana,$^{1,2,}\footnote{mfontana11@uninsubria.it}$ \hspace{0.5mm} and Massimiliano Rinaldi$^{3,4,} \footnote{massimiliano.rinaldi@unitn.it}$}
\date{\vspace{-9ex}}

\numberwithin{equation}{section}
\numberwithin{figure}{section}

\begin{document}
\maketitle

\begin{center}
    $^{1}$ \emph{Dipartimento di Scienza e Alta Tecnologia, Università degli Studi dell’Insubria, \\
    Via Valleggio 11, I-22100, Como, Italy} \\
    \vspace{0.3cm}
    $^{2}$ \emph{INFN, Sezione di Milano, Via Celoria 16, 20133, Milano, Italy} \\
    \vspace{0.3cm}
    $^{3}$ \emph{Dipartimento di Fisica, Università di Trento, \\
      Via Sommarive 14, I-38123 Povo (TN), Italy} \\ 
    \vspace{0.3cm}
    $^{4}$ \emph{Trento Institute for Fundamental Physics and Applications TIFPA-INFN \\
      Via Sommarive 14, I-38123 Povo (TN), Italy}
    \vspace{0.5cm}
\end{center}

\abstract{
    Hawking radiation can be regarded as a spontaneous and continuous creation of
virtual particle-antiparticle pairs outside the event horizon of a black hole where strong
tidal forces prevent the annihilation: the particle escapes to infinity contributing to
the Hawking flux, while its corresponding antiparticle partner enters the event horizon
and ultimately reaches the singularity. The aim of this paper is to investigate the
energy density correlations between the Hawking particles and their partners across the
event horizon of two models of non-singular black holes by calculating the two-point correlation
function of the density operator of a massless scalar field. This analysis is motivated by the fact that in acoustic
black holes particle-partner correlations are signaled by the presence of a peak in the
equal time density-density correlator. Performing the calculation in a Schwarzschild black hole it was shown in \cite{ArticoloPrincipale} that
the peak does not appear, mainly because of the singularity. It is
then interesting to consider what happens when the singularity is not present. In the
Hayward and Simpson-Visser non-singular black holes we show that the density-density correlator
remains finite when the partner particle approaches the hypersurface that replaces
the singularity, opening the possibility that partner-particle correlations can propagate
towards other regions of spacetime instead of being lost in a singularity.}

\newpage

\section{Introduction}

One of the most important results of modern theoretical physics is the fact that black holes are not completely black, as predicted by general relativity, but they emit thermal radiation. This process was demonstrated by Hawking \cite{Hawking:paper} in the context of quantum field theory in curved spacetime and it is caused by the dynamical gravitational field that acts during the collapse of a massive astrophysical body leading to the formation of a black hole. Even though the region inside the event horizon is causally disconnected from the rest of the spacetime, analogue models of gravity have shown that the equal time density-density correlator can be used to study the Hawking process when one point is taken outside the event horizon and the other inside. These models are based on the pioneering work of Unruh \cite{Unruh}, who established that there exists an analogous phenomenon to Hawking radiation in condensed matter systems, where sound waves play the role of light and sound horizons the role of event horizons. This opened the possibility to study the Hawking radiation process in completely different physical systems. As an example, the behavior of a massless scalar field on a curved background is explained by the same equation that characterizes the propagation of sound waves in Eulerian fluids, which are described in terms of an acoustic metric that is a function of the background flow. In particular, the curvature of the acoustic geometry is induced by the inhomogeneities in the fluid flow, while flat Minkowski spacetime is recovered in the case of a homogeneous system. It is also possible to simulate a black hole using what is called an acoustic black hole, which is obtained whenever a subsonic flow is turned supersonic: sound waves in the supersonic region are dragged away by the flow and cannot propagate back toward the acoustic horizon separating the supersonic and subsonic regions \cite{AcousticCorrelations}. However, differently from the case of astrophysical black holes, both the external and internal regions are accessible to experiments. In the case of acoustic black holes constructed from Bose-Einstein condensates, it has been predicted that the correlations between the Hawking particles and their partners will form a stationary peak in the equal time density-density correlator, which appears at late times after the formation of the sonic horizon, with one point taken inside it and the other outside \cite{AcousticCorrelations}, \cite{RampupB&F}. This striking feature has indeed been experimentally observed \cite{Steinhauer}, \cite{Nova} and it is the most stringent evidence of Hawking-like (phonons in this case) radiation in an analogue black hole. Inspired by these results, R. Balbinot and A. Fabbri \cite{ArticoloPrincipale} calculated the equal time density-density correlator of a massless scalar field on a Schwarzschild black hole background, finding a result which is in disagreement with the acoustic black hole case: the expected peak signaling the particle-partner correlations does not appear. As we will review in detail below,  the reason lies in the fact that when a Hawking particle emerges out of vacuum fluctuations in a region outside the event horizon, called quantum atmosphere, the corresponding partner has already entered the singularity. Hence the peak does not have sufficient time to form. On the opposite, in the acoustic case, the singularity does not exist and one can tune the experiment setup in such a way that the peak can always form.

The aim of this paper is to study the correlations in regular spacetimes by calculating the density-density correlator in the Hayward \cite{Hayward:NSBH} and Simpson-Visser \cite{SV:NSBH} non-singular black holes. 
This choice is motivated by the fact that the latter represents a minimal modification to the Schwarzschild black hole that makes it regular, while the former possesses two horizons, of which the inner one is a Cauchy horizon, and therefore it has a causal structure that is similar to the usual charged and rotating solutions of general relativity. 
The result is that unlike in the Schwarzschild case, where the equal time density-density correlator vanishes when the partner particle enters the singularity, in the regular black holes analyzed here the correlator remains finite when the partner particle approaches the hypersurfaces that replace the singularity. \newline
The paper is organized as follows.
In Section \ref{hawkrad} we briefly recall the fundamental concepts of Hawking radiation. In Section \ref{sec.3} we review the results concerning a Schwarzschild black hole. In Section \ref{sec.4} we generalize the procedure to discuss possible correlations in the Hayward and Simpson-Visser non-singular black holes. Section \ref{sec.5} is devoted to the conclusions. All the mathematical details can be found in the appendices \ref{Appendix A}, \ref{Appendix B}, \ref{Appendix C}, \ref{Appendix D}, \ref{Appendix E}.

\section{Hawking radiation}\label{hawkrad}
We now briefly recall some fundamental aspects of Hawking radiation that will be useful in the rest of the paper. Let us consider a massless scalar field propagating in a black hole spacetime, which possesses asymptotically stationary regions in the past ("in"), corresponding to past null infinity, and in the future ("out"), given by future null infinity. Stationarity implies the existence of a timelike Killing vector field with respect to which one can uniquely specify positive frequency mode solutions to the field equation and the corresponding vacuum states, defined as the absence of particles according to all inertial observers in the asymptotic region of interest\footnote{Note that, in principle, it is not possible to define a complete set of positive frequency mode solutions at future null infinity ($\mathscr{I^{+}}$) since the latter is not a proper Cauchy surface. To form a complete Cauchy surface it is necessary to consider the union of $\mathscr{I^{+}}$ with the event horizon so that among the outgoing modes we must distinguish between the ones that are able to reach $\mathscr{I^{+}}$ and the ones that are trapped inside the horizon. However, an explicit expression of the latter, which would be difficult to obtain because there is no natural choice of time at the horizon, is not needed to evaluate the particle production at $\mathscr{I^{+}}$ \cite{Fabbri:book}.}. Let $\ket{in}$ be the vacuum state in the "in" region and $\ket{out}$ the one in the "out" region. One usually works in the Heisenberg picture, so that by assuming that the quantum state of the field in the "in" region is $\ket{in}$, it will remain in that state during its subsequent evolution. However, as it was first shown by Hawking \cite{Hawking:paper}, inertial observers in the "out" region will detect a thermal distribution of particles at the temperature $T_{H}=\frac{k}{2\pi}$, where $k$ is the surface gravity of the event horizon. One then concludes that particles have been created by the external time-dependent gravitational field acting between the two asymptotic stationary regions. In fact, 
 by considering the field propagation in a Schwarzschild background, it is possible to write the $\ket{in}$ vacuum state in the "out" region formally as \cite{Fabbri:book}
\begin{equation}
    \ket{in}\propto e^{-\sum_{\omega}\frac{\pi \omega}{k}a_{\omega}^{bh\dagger}a_{\omega}^{out\dagger}}\ket{out},
\end{equation}
where $a_{\omega}^{out\dagger}$ and $a_{\omega}^{bh\dagger}$ are the creation operators for respectively the outgoing modes reaching future null infinity and the trapped modes entering the horizon. The mathematical expressions of the states on which these operators act depend on the choice of the orthonormal set of modes that are exact solutions of the Klein-Gordon equation governing the excitation (usually a massless scalar field) \cite{,Fabbri:book,Brout:1995rd}. In this formalism, the $\ket{in}$ vacuum state represents a flux of entangled particles, one escaping to infinity and the other crossing the black hole horizon. This allows an intuitive picture of the Hawking process: the presence of a trapped region acts as an energy reservoir for the continuous and spontaneous creation of particle-antiparticle pairs outside of the event horizon, where strong tidal forces prevent their mutual annihilation. The particle, having positive Killing energy, escapes to infinity contributing to the Hawking flux, while its corresponding antiparticle partner enters the event horizon and ultimately reaches the singularity, depleting the trapped region due to its negative Killing energy.

\section{Quantum correlations in the Schwarzschild black hole}
\label{sec.3}
\noindent
In this section we review the calculation of the energy density correlations between Hawking quanta across the event horizon of a Schwarzschild black hole \cite{ArticoloPrincipale}. 

\subsection{Modelling gravitational collapse}
Hawking radiation is produced by the time-dependent gravitational field during the collapse that leads to the formation of a black hole. In the spirit of the "no hair" theorem \cite{nohairtheorem}, the final result should be insensitive to the details of the collapse and thus one can work with the simplest solution to Einstein's equation describing the formation of a black hole through gravitational collapse, namely the Vaidya metric. This is obtained by expanding the mass parameter in the Schwarzschild metric from a constant to a function of the ingoing Eddington-Finkelstein coordinate $v$:
\begin{equation}
\label{eq:3.1}
d s^{2}=-\left(1-\frac{2 M(v)}{r}\right) d v^{2}+2dvdr+r^{2} d \Omega^{2}.
\end{equation}
Since the Ricci tensor has only one non-vanishing component given by
\begin{equation}
    R_{vv}=\frac{2}{r^{2}}\frac{dM(v)}{dv}.
\end{equation}
and the Ricci scalar is zero, the only non-vanishing component of the stress-energy tensor is
\begin{equation}
    T_{vv}=\frac{L(v)}{4\pi r^{2}},
\end{equation}
where $L(v)=\frac{dM(v)}{dv}$.
The physical interpretation is that the Vaidya solution describes a purely ingoing flux of massless radiation characterized by the function $L(v)$. If such influx is turned on at some advanced time $v_{i}$ and turned off at $v_{f}$, then the spacetime geometry can be  divided into three regions:
\begin{itemize}
    \item A Minkowski vacuum region $v<v_{i}$;
    \item An intermediate collapse region $v_{i}<v<v_{f}$;
    \item The final Schwarzschild configuration $v>v_{f}$.
\end{itemize}
To discuss the Hawking radiation one cares about the "in" and "out" stationary regions, so we can ideally narrow the collapse region down to a single null surface. Therefore, we consider an ingoing shock wave located at some $v=v_{0}$ of the form $L(v)=M\delta (v-v_{0})$, that is $M(v)=M\theta (v-v_{0})$.
The resulting spacetime is then obtained by patching portions of Minkowski and Schwarzschild spacetimes along $v=v_{0}$.
For $v<v_{0}$ the metric is Minkowskian and can be written in double null form:
\begin{equation}
    ds^{2}=-du_{in}dv+r^{2}(u_{in},v)d\Omega ^{2},
\end{equation}
where $u_{in}=t-r=v-2r$.
For $v>v_{0}$ the metric is the Schwarzschild one describing a black hole of mass $M$ and horizon located at $r=2M$. In double null form it reads
\begin{equation}
    ds^{2}=-\left(1-\frac{2M}{r}\right) dudv+r^{2}(u,v)d\Omega ^{2},
\end{equation}
where
\begin{equation}
\label{eq:3.6}
    u=t-r^{*}=v-2r^{*},
\end{equation}
and $r^{*}=\int \left(1-\frac{2M}{r}\right)^{-1}dr$ is the tortoise coordinate.\\
To guarantee continuity of the global metric at $v=v_{0}$, one needs to impose the condition
\begin{equation}
\label{eq:3.7}
    r(u_{in},v_{0})=r(u,v_{0}),
\end{equation}
which leads to the following relation between the retarded null coordinates inside and outside $v_{0}$:
\begin{equation}
\label{eq:3.8}
    u=u_{in}-4M\ln{\abs{\frac{u_{in}}{4M}}}.
\end{equation}
Exploiting the arbitrariness of $v_{0}$, we have set $v_{0}=4M$ to simplify calculations. Inverting \eqref{eq:3.8} one can formally extend the coordinate $u_{in}$ in the exterior region in terms of the Lambert function \cite{Lambert}
\begin{equation}
\label{eq:3.9}
    u_{in}=-4MW\left(\pm e^{-\frac{u}{4M}}\right).
\end{equation}
The positive sign holds in the exterior region and the minus sign in the interior one.
\newline

\subsection{Density-density correlator}
\label{The density-density correlator}
Let us now consider a massless scalar field propagating in the Vaidya spacetime. We assume the quantum state of the field to be the Minkowski vacuum $\ket{in}$ at past null infinity. Then, we neglect the backscattering of the modes induced by the curvature of spacetime, impose reflecting boundary conditions at the origin $r=0$ in the Minkowski region, and require regularity of the modes there. This corresponds to work on the effective (1+1)  metric \cite{Fabbri:book}, \cite{Birrell&Davies:book}, \cite{Parker&Toms:book}
\begin{equation}
\label{eq:3.10}
    d s_{(2)}^{2}=-\left(1-\frac{2 M(v)}{r}\right) d v^{2}+2dvdr.
\end{equation}
As shown in Appendix \ref{Appendix B}, the density-density correlator of the scalar field is given by the action of a differential operator on the Wightman function, which is defined as
\begin{equation}
\label{eq:3.11}
    G^{+}(x,x^{\prime})=\bra{in}\phi(x)\phi(x^{\prime})\ket{in}.
\end{equation}
$G^{+}(x,x^{\prime})$ can be easily computed by expanding the field on the complete set of positive frequency mode solutions to the field equation at past null infinity \cite{Birrell&Davies:book}, namely
\begin{equation}
\label{eq:3.12}
    G^{+}(x,x^{\prime})=-\frac{1}{4 \pi} \ln \frac{\left(u_{i n}-u_{i n}^{\prime}\right)\left(v-v^{\prime}\right)}{\left(u_{i n}-v^{\prime}\right)\left(v-u_{i n}^{\prime}\right)}.
\end{equation}
As stressed in \cite{ArticoloPrincipale}, to fully characterize the correlation functions, one needs to specify the observer's state and we choose the free-falling one on a radial path with velocity vector $u^{\alpha}$. Due to the conformal flatness of the metric \eqref{eq:3.10}, the energy density of the field $\phi$ is equivalent to its pressure density measured by this class of observers. Thus (for more details see Appendix \ref{Appendix A})
\begin{equation}
\label{eq:3.13}
    \rho=T_{\alpha\beta}u^{\alpha}u^{\beta},
\end{equation}
where $T_{\alpha\beta}$ is the stress-energy tensor of a massless scalar field,
\begin{equation}
\label{eq:3.14}
    T_{\alpha\beta}=\partial_{\alpha}\phi \partial_{\beta}\phi - \frac{g_{\alpha\beta}}{2}g^{\mu\nu}\partial_{\mu}\phi\partial_{\nu}\phi.
\end{equation}
To obtain a simple expression for the energy density it is convenient to map the metric \eqref{eq:3.10} into the "out" region of the Painlevé-Gullstrand coordinates \cite{GPcoordinates}
\begin{equation}
\label{eq:3.15}
    ds_{(2)}^{2}=-fdT^{2}-2VdTdr+dr^{2},
\end{equation}
where $f=1-\frac{2M}{r}$, $V=-\sqrt{\frac{2M}{r}}$ and $T$ is the Painlevé time given by
\begin{equation}
\label{eq:3.16}
    T=v+\int\left(\frac{\sqrt{1-f}-1}{f}\right)dr.
\end{equation}
As also shown in Appendix \ref{Appendix A}, the energy density measured by radial free-falling observers in Painlevé-Gullstrand coordinates is 
\begin{equation}
    \rho=T_{rr}.
\end{equation}
Writing this expression in double null coordinates allows us to obtain the following result for the density-density correlator (see Appendix \ref{Appendix B} for details):
\begin{equation}
\label{eq:3.18}
\begin{aligned}
G\left(x, x^{\prime}\right) &=\bra{in}\rho(x)\rho(x^{\prime})\ket{in}=\\ &=\bra{in}\frac{T_{uu}(x)T_{u^{\prime}u^{\prime}}\left(x^{\prime}\right)}{(1+V(x))^{2}\left(1+V\left(x^{\prime}\right)\right)^{2}}+\frac{T_{uu}(x)T_{v^{\prime}v^{\prime}}\left(x^{\prime}\right)}{(1+V(x))^{2}\left(1-V\left(x^{\prime}\right)\right)^{2}}+ \\
&+\frac{T_{vv}(x)T_{u^{\prime}u^{\prime}}\left(x^{\prime}\right)}{(1-V(x))^{2}\left(1+V\left(x^{\prime}\right)\right)^{2}}+\frac{T_{vv}(x)T_{v^{\prime}v^{\prime}}\left(x^{\prime}\right)}{(1-V(x))^{2}\left(1-V\left(x^{\prime}\right)\right)^{2}}\ket{in}.
\end{aligned}
\end{equation}

\begin{figure}
\centering
\includegraphics[width=10cm, height=6cm]{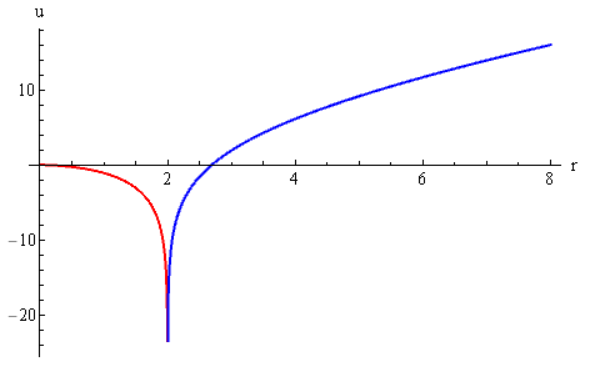}
\caption{Plot of the left hand side of equation \eqref{eq:3.22} ($u^{\prime})$ as a function of $r^{\prime}<2M$ in red and of the right hand side ($u$) for $r>2M$ in blue.}
\label{fig:3.4}
\end{figure}

As pointed out in \cite{Fabbri:book},  (see also \cite{Candelas}, \cite{C&F}, \cite{stresstensorinRN}), the expectation value of a physical observable in the $\ket{in}$ vacuum state which describes gravitational collapse (and which is reproduced at late time by the Unruh state in an eternal Schwarzschild black hole) can be written as the sum of two contributions, one describing the particle creation induced by the formation of a black hole and the other being the vacuum polarization that is constant in time and does not contribute to the Hawking radiation. The vacuum polarization terms come in fact from the expectation value of the stress-energy tensor computed in the Boulware state, which is chosen to correspond to the usual Minkowski vacuum at future null infinity. However, this vacuum state is pathological at the horizon, in the sense that the expectation values of physically relevant quantities diverge as the horizon is approached. Being empty at infinity, the Boulware state corresponds to the absence from the vacuum of black body radiation at the black hole temperature and so its physical realization would be the state describing vacuum polarization outside a massive spherical body of radius slightly larger than its Schwarzschild radius. Since we are interested in studying energy density correlations between Hawking quanta we therefore have to neglect vacuum polarization effects that do not describe black holes evaporation. As shown in \cite{ArticoloPrincipale}, \cite{RN}, this accounts for retaining only the term coming from the $u$ sector of the correlator, since in a two-dimensional spacetime all the other terms receive contributions only from the vacuum polarization. Therefore, the key ingredient for the description of density correlations is the $T_{uu}$ part, which, as reported in Appendix \ref{Appendix B}, is obtained from 

\begin{align}
\label{eq:3.19}
    &\frac{1}{(1+V(r))}\frac{1}{(1+V(r^{\prime}))}\partial_{u}\partial_{u^{\prime}}\bra{in}\phi(x)\phi(x^{\prime})\ket{in}_{|T=T^{\prime}}=\notag\\
    &=\left.\frac{1}{16\pi}\frac{1}{1-\sqrt{\frac{2M}{r}}}\frac{1}{1-\sqrt{\frac{2M}{r^{\prime}}}}\frac{1}{(u_{in}-4M)(u^{\prime}_{in}-4M)}\frac{1}{\cosh^{2}{\ln{\sqrt{-\frac{u_{in}}{u^{\prime}_{in}}}}}}\right|_{T=T^{\prime}}.
\end{align}
At late retarded time ($u\to+\infty, u_{in}\to0$) the relation between the retarded null coordinates inside and outside the horizon reported in equation \eqref{eq:3.9} can be approximated as 
\begin{equation}
\label{eq:3.20}
    u_{in}\simeq\pm4Me^{-\frac{u}{4M}},
\end{equation}
where $u_{in}<0$ for $r>2M$ and $u_{in}>0$ for $r<2M$.
In this limit one obtains 
\begin{align}
\label{eq:3.21}
    &\frac{1}{(1+V(r))}\frac{1}{(1+V(r^{\prime}))}\partial_{u}\partial_{u^{\prime}}\bra{in}\phi(x)\phi(x^{\prime})\ket{in}_{|T=T^{\prime}}\simeq\notag\\
    &\left.\simeq\frac{1}{16\pi}\frac{1}{1-\sqrt{\frac{2M}{r}}}\frac{1}{1-\sqrt{\frac{2M}{r^{\prime}}}}\frac{1}{16M^{2}}\frac{1}{\cosh^{2}{\left(\frac{u-u^{\prime}}{8M}\right)}}\right|_{T=T^{\prime}}.
\end{align}
This function has a maximum for $u=u^{\prime}$, i.e. along the trajectory of the Hawking particle and its partner, as in the case of analogue black holes. However, one has still to verify whether this maximum condition is fulfilled or not. Considering the relation between Eddington-Finkelstein coordinates and Painlevé time \eqref{eq:3.16}, the condition $u=u^{\prime}$ at equal times can be written as
\begin{equation}
\label{eq:3.22}
    r+2\sqrt{2Mr}+4M\ln{\left(\sqrt{\frac{r}{2M}}-1\right)}=r^{\prime}+2\sqrt{2Mr^{\prime}}+4M\ln{\left(1-\sqrt{\frac{r^{\prime}}{2M}}\right)}.
\end{equation}

In analogy to what happens for acoustic black holes, one would expect that at late times and for $r$ sufficiently far away from the horizon, the equal time density-density correlator should show a peak along \eqref{eq:3.22}, see \cite{ArticoloPrincipale,AcousticCorrelations,RampupB&F}. \\
The key ingredient to investigate this possibility is the location of the region of spacetime where Hawking quanta emerge out of vacuum fluctuations. \\
Calculating the radius of a radiating body using the Stephan-Boltzmann law, Giddings \cite{Giddings:QA} was able to show that the Hawking particles originate from a region outside the event horizon of the black hole, called "quantum atmosphere", located at a distance $o(1/k)$, $k$ being the horizon surface gravity. This result was later corroborated by the work of other authors \cite{Liberati:QA}, \cite{QA2}, using a detailed analysis of the renormalized vacuum expectation value of the stress-energy tensor of a massless scalar field in the Schwarzschild spacetime. The same conclusion was obtained also for acoustic black holes \cite{RampupB&F}, \cite{ArticoloPrincipale}. \\
If we now plot the two functions entering the left and right-hand sides of equation \eqref{eq:3.22} (Fig. \ref{fig:3.4}) we see that the right-hand side is always smaller or equal than zero since $0<r^{\prime}<2M$. As a consequence, to find a solution, also the left-hand side has to be smaller or equal than zero and this happens for $r\lesssim2.6M$. Therefore, equation \eqref{eq:3.22} has real solutions only if $r\lesssim2.6M$, for which the corresponding $r^{\prime}$ is located between zero and $2M$. We then conclude that when the Hawking particle emerges from the quantum atmosphere out of vacuum fluctuations at a distance of the order of $4M$ from the horizon, the corresponding partner has already been swallowed by the singularity and the correlations are lost. On the other hand, for solutions with a non-vanishing $r^{\prime}<2M$, the correlator $G\left(x, x^{\prime}\right)$ is dominated by the coincidence limit and the peak, signaling the particle-partner correlations, does not appear. \\
This result is in striking disagreement with what happens in acoustic black holes and it is due to the presence of the quantum atmosphere and of the central singularity.

For later comparison with other black hole type curved spacetime backgrounds, as well as to better understand the behavior of the correlator in regions of physical interest, it is useful to study the correlator in the limits $r^{\prime}\to 0$, $r\to 2M$ and $r\to\infty$. Using the results of Appendix \ref{Appendix B}, the $T_{uu}$ correlator can be written as
\begin{equation}
\label{eq:3.23}
    G\left(x,x^{\prime}\right)=\frac{1}{16\pi^{2}}\frac{1}{\left(1+V\left(x\right)\right)^{2}\left(1+V\left(x^{\prime}\right)\right)^{2}}\left(\frac{du_{in}}{du}\right)^{2}\left(\frac{du^{\prime}_{in}}{du^{\prime}}\right)^{2}\frac{1}{(u_{in}-u^{\prime}_{in})^{4}}.
\end{equation}
\newline
We note that fixing the time variables, \eqref{eq:3.23} takes the form of the product of a function of $r$, a function of $r^{\prime}$ and another of both $r$ and $r^{\prime}$. However, the latter affects the form of the correlator only in the coincidence limit or when the denominator diverges, which occurs when one of the two particles is at infinity. All the mathematical details regarding the calculation of the limits of the correlator are described in appendix \ref{Appendix C}.\newline
When $r^{\prime}\to 0$ we can focus on the parts that depend only on $r^{\prime}$:
\begin{equation}
    \lim_{r^{\prime}\to 0}\frac{1}{(1\pm V(r^{\prime}))^{2}}=0,
\end{equation}
and 
\begin{equation}
    \lim_{r^{\prime}\to 0}\frac{du^{\prime}_{in}}{du^{\prime}}=\infty.
\end{equation}
Thus, one needs to consider
\begin{equation}
\label{eq:3.26}
    \lim_{r^{\prime}\to 0}\frac{1}{(1+V(r^{\prime}))^{2}}\left(\frac{du^{\prime}_{in}}{du^{\prime}}\right)^{2}=0.
\end{equation}
Therefore, when the partner particle approaches the singularity, the density-density correlator vanishes.
\newline
When $r\to 2M$ one has 
\begin{equation}
    \lim_{r\to 2M}\frac{1}{(1+V(r))^{2}}=\infty,
\end{equation}
\begin{equation}
    \lim_{r\to 2M}\frac{du_{in}}{du}=0.
\end{equation}
In this case we have to study the following limit
\begin{equation}
\label{eq:3.29}
    \lim_{r\to 2M}\frac{1}{(1+V(r))^{2}}\left(\frac{du_{in}}{du}\right)^{2}=4e^{-\frac{T}{2M}}.
\end{equation}
In equation \eqref{eq:3.29}, $T$ is the Painlevé time it takes to go from the quantum atmosphere $(r_{qa}\sim\frac{1}{k}=4M)$ to the horizon $(r=2M)$:
\begin{equation}
    T=\int_{4M}^{2M}\frac{1}{V}dr=\frac{4}{3}(2\sqrt{2}-1)M.
\end{equation}
The same result is obtained for $r^{\prime}\to 2M$. Therefore, the correlator remains finite when one of the Hawking quanta approaches the horizon.
\newline
When $r\to\infty$, the part of the correlator that depends on both $r$ and $r^{\prime}$ dominates because $u_{in}\to u\to \infty$. This implies that the correlator goes to zero when the Hawking particle reaches spatial infinity.

\section{Quantum correlations in non-singular black holes}
\label{sec.4}
As discussed in the previous section, energy density correlations between Hawking particles and their partners across the event horizon of a Schwarzschild black hole are lost because of the presence of the quantum atmosphere and of the singularity. It would then be interesting to study what happens if the singularity is not present.
\noindent
\subsection{Non-singular black holes}
One of the fundamental results of classical general relativity is the Penrose singularity theorem \cite{Penrose}, which proves that, under general assumptions, the gravitational collapse of a sufficiently massive astrophysical body will lead to the formation of trapped surfaces and thus of a singularity. However, the appearance of singularities is usually accompanied by an unlimited increase of spacetime curvature, and in these conditions, Einstein's equation is not expected to work anymore because the quantum corrections become of the same order of the main classical terms. Therefore, singularities in general relativity are usually regarded as a problem of this classical theory, which may be solved by its quantization. In the absence of a completely satisfactory theory of quantum gravity, one can still provide qualitative arguments for the existence of non-singular black holes, violating at least one of the assumptions of the Penrose theorem, typically the strong energy condition \cite{NSBHreview}, \cite{notesnsbhs}.
One possibility is to study quantum effects of gravitation with the semiclassical approach in which matter fields are quantized in the usual way, while the spacetime geometry is treated classically. However, in the case of black hole spacetimes, close to the singularity the curvature reaches order unity in Planck units and so quantum fields begin to dominate on the geometry. Whether gravitation can still be treated classically at this level is far from being certain, but one can wonder if quantum polarization effects can provide a mechanism to slow down the infinite rise of curvature and to maintain it bounded to Planckian magnitude \cite{vacuumnsbh}.
It is possible to outline in a qualitative way the behavior of the corrected curvature when quantum effects are taken into account assuming that the $\langle -T^{t}_{t}\rangle$ component of the stress-energy tensor is proportional to the curvature squared\footnote{Here by curvature squared we intend the Kretschmann scalar, which for the Schwarzschild black hole is $K\sim\frac{M^{2}}{r^{6}}$.} with a coefficient $(a^{2})$, that depends on the number and on the nature of the quantum fields \cite{Poisson&Israel}. One then arrives at the following expression:
\begin{equation}
\label{eq:4.1}
    \frac{M(r)}{r^{3}}=\frac{1}{a^{2}+\left(\frac{r}{r_{Q}}\right)},
\end{equation}
where $M(r)$ is the mass function and $r_{Q}=M^{1/3}$ is the radius at which quantum effects become important. For $r\gg  r_{Q}$ one recovers the curvature of the Schwarzschild black hole, while for smaller radii it depends on the sign of $a^{2}$. If $a^{2}<0$ the curvature diverges at $r=|a^{2/3}|r_{Q}$, while if $a^{2}>0$ the curvature remains bounded and constant. Therefore, a spherically symmetric, uncharged black hole should be described by the Schwarzschild solution down to a critical radius $r_{Q}$ where quantum effects produce a smooth transition towards a constant curvature, de Sitter core \cite{Balbinot&Poisson}. In this region, the strong energy condition does not hold and thus the singularity can be avoided. \newline
It is interesting to notice that the same conclusion was obtained also in \cite{Markov}, where it was supposed that, because of quantum effects, the spacetime curvature should always be subject to an upper bound of Planckian magnitude. At the limiting curvature, the density is so high that all the particles lose their identity and matter undergoes a transition into a vacuum-like state described by the stress-energy tensor $T_{\mu\nu}=-\rho g_{\mu\nu}$, where $\rho$ is a positive, constant energy density. In this way, the strong energy condition is violated and gravity acts in such a way that the trajectories of freely falling test particles moving along causal geodesics behave as if they were repulsed from the center.
In the following, we consider a quantized scalar field that propagates on the non-singular geometries of the Simpson-Visser and Hayward black holes. As a first approximation, we neglect the back-reaction of the field on the background geometry. As a consequence, the stress-energy tensor associated to the scalar field cannot contribute to energy condition violations.

\subsubsection{Hayward non-singular black hole}
We now present an explicit example of a non-singular black hole obtained by applying the argument presented above. As previously discussed, assuming that the $\langle -T^{t}_{t}\rangle$ component of the stress-energy tensor is proportional to the curvature squared because of quantum polarization effects, the mass function can be written as (see eq. \eqref{eq:4.1})
\begin{equation}
    M(r)=\frac{Mr^{3}}{r^{3}+2ML^{2}},
\end{equation}
where $2L^{2}=a^{2}>0$. It is then possible to consider the following spherically symmetric metric:
\begin{equation}
\label{eq:4.3}
    ds^{2}=-f(r)dt^{2}+f(r)^{-1}dr^{2}+r^{2}d\Omega^{2},
\end{equation}
where
\begin{equation}
\label{eq:4.4}
    f(r)=1-\frac{2M(r)}{r}=1-\frac{2Mr^{2}}{r^{3}+2L^{2}M}.
\end{equation}
This is called the Hayward metric \cite{Hayward:NSBH}. \newline
Asymptotically, for $r\gg M$, $f(r)$ behaves as 
\begin{equation}
    f(r)\sim 1-\frac{2M}{r}, 
\end{equation}
reproducing a Schwarzschild spacetime with mass $M$. \newline
For $r\to 0$, we have
\begin{equation}
    f(r)\sim 1-\frac{r^{2}}{L^{2}},
\end{equation}
which gives a de Sitter spacetime with cosmological constant $\Lambda=\frac{3}{L^{2}}$. Hayward's model is the most popular of a class of black holes that have a de Sitter structure near the central singularity. In the literature, these are often called non-singular black holes with a de Sitter core (for a recent review see \cite{Sebastiani:2022wbz}).  \newline
It is possible to check that the Hayward metric is not singular by computing the curvature invariants, which are regular everywhere, and also that the stress-energy tensor that generates such a spacetime violates the strong energy condition.
\newline
The Hayward spacetime contains a trapped region, the boundary of which is given by the solutions of the equation $f(r)=0$.
In particular, it is possible to define a critical mass $M^{*}=\frac{3\sqrt{3}L}{4}$ such that, if $M>M^{*}$, $f(r)$ has three real roots, if $M=M^{*}$, $f(r)$ has one multiple root, if $M<M^{*}$, $f(r)$ has one real root and a complex pair. However, in the $M>M^{*}$ case one of the three roots is negative and thus it cannot be accepted because $r$ should be greater or equal then zero (and real). The same happens for the real root in the $M<M^{*}$ case. Since $L$ gives the approximate length scale below which quantum effects of gravity become dominant, one might expect $L$ to be of the order of the Planck length. Therefore, for black holes formed by gravitational collapse, $M$ is much greater than the critical mass $M^{*}$\footnote{This is the case for the calculation of the density-density correlator, that is when the geometry has already settled down after the collapse and the backreaction of Hawking evaporation on the metric can still be neglected.}, and the roots of $f(r)$ can be approximated as \cite{Hayward:NSBH}
\begin{equation}
\label{root1}
    r_{1}\simeq 2M-\frac{L^{2}}{2M},
\end{equation}
\begin{equation}
\label{root2}
    r_{2}\simeq L+\frac{L^{2}}{4M}.
\end{equation}
It is important to note that the Hayward spacetime is not globally hyperbolic because the inner horizon at $r=r_{2}$ is a Cauchy horizon. This is a common characteristic of black holes with a de Sitter core. \\
The Hayward non-singular black hole can be obtained from gravitational collapse generalizing the Vaidya metric used in the previous section to model the formation of a Schwarzschild black hole. Also in this case, it is enough to promote the mass parameter $M$ in the metric \eqref{eq:4.3} from a constant to a function of the ingoing null coordinate $v$ \cite{Hayward:NSBH}.

\subsubsection{Simpson-Visser non-singular black hole}
Another popular model of regular spacetime is the one proposed by Simpson and Visser \cite{SV:NSBH}. In contrast with the case studied above, this model does not have a de Sitter core \cite{Sebastiani:2022wbz}. This is an example of "black bounce" spacetime \cite{BlackBounce}, which is a regular spacetime, where the area radius always remains non-zero, thereby leading to a "throat". In the case that we are going to analyze, the idea is to introduce a minimal modification to the Schwarzschild metric in order to make it regular. The line element can be written as
\begin{equation}
\label{eq:4.9}
    ds^{2}=-\left(1-\frac{2M}{\sqrt{r^{2}+a^{2}}}\right)dt^{2}+\frac{dr^{2}}{\left(1-\frac{2M}{\sqrt{r^{2}+a^{2}}}\right)}+(r^{2}+a^{2})(d\theta^{2}+\sin^{2}{\theta}d\varphi^{2}).
\end{equation}
Depending on the value of the parameter $a$, this metric represents either:
\begin{itemize}
    \item The ordinary Schwarzschild spacetime $(a=0)$;
    \item A traversable wormhole geometry (in the Morris-Thorne sense \cite{Thorne}) with a two-way timelike throat $(a>2M)$;
    \item A one-way wormhole geometry with an extremal null throat $(a=2M)$;
    \item A non-singular black hole $(a<2M)$.
\end{itemize}
In particular, we will be interested in the latter case. \newline
The horizons locations are obtained by solving the equation $f(r)=0$, with $f(r)=1-\frac{2M}{\sqrt{r^{2}+a^{2}}}$. One has $r_{\pm}=\pm\sqrt{4M^{2}-a^{2}}$, which are real, non-degenerate, solutions if and only if $a<2M$, corresponding to the case under study. Therefore, when $a<2M$ there will be symmetrically placed $r$ coordinate values, $r_{+}>0$ and $r_{-}<0$, which correspond to a pair of horizons. \newline
The hypersurface $r=0$ is a spacelike, spherical hypersurface which marks the boundary between our universe and a bounce into a separate copy of it. This implies that for negative values of the $r$ coordinate we have "bounced" into another universe. 
It is possible to compute the curvature invariants to directly check the regularity of the spacetime. Then, studying the stress-energy tensor that gives rise to such a spacetime, one can show that the null energy condition is violated, and so are all the energy conditions.
\newline
A peculiar feature of the Simpson-Visser non-singular black hole is the fact that it is not possible to describe its formation from gravitational collapse generalizing the Vaidya metric if one keeps the parameter $a$ constant and non-vanishing because it implies the existence of a throat. In fact, it can be easily seen from the metric \eqref{eq:4.9} that when $M=0$ one has a wormhole instead of flat space. Therefore, starting with zero mass and then considering an increasing mass function one has a transition from a wormhole to a non-singular black hole \cite{SV:Vaidya}. This implies that a more complete model of gravitational collapse is needed to describe the onset of a black bounce. However, this is not a problem for the calculation of the density-density correlator, which is the focus of this work.

\subsection{Density-density correlator}
To study the density-density correlator in non-singular black hole spacetimes we follow the procedure adopted in Section \ref{sec.3}, where we performed the same calculation on a Schwarzschild background spacetime. In particular, we neglect backscattering of the modes induced by the curvature of spacetime and consider a $(1+1)$ dimensional theory describing the propagation of a massless scalar field in the $(1+1)$ dimensional section of the Vaidya-like spacetime described by the line element
\begin{equation}
\label{eq:4.10}
    ds^{2}_{(2)}=-f(r,v)dv^{2}+2dvdr.
\end{equation}
In this way, the Wightman function is the same as \eqref{eq:3.12} because any two-dimensional metric describing the "in" region is conformal to the Minkowski one, and therefore the field equation has the same mode solutions at past null infinity. Then, we consider the quantum state of the system to be the $\ket{in}$ vacuum of the "in" region. Therefore, the correlator that we want to study is 
\begin{equation}
    G(x,x^{\prime})=\bra{in}\rho(x)\rho(x^{\prime})\ket{in},
\end{equation}
evaluated in the black hole region, where one point $(x)$ is taken outside the (outer) horizon and the other $(x^{\prime})$ inside. It is also assumed that the spacetime has already settled down after the collapse (or the wormhole to black hole transition in the Simpson-Visser case) and that the backreaction of Hawking evaporation on the metric can be neglected. In this way, one has to deal with static, spherically symmetric metrics for which it is possible to introduce Painlevé coordinates, describing free-falling observers, in the same way as in Schwarzschild, and write the metric in the form \eqref{eq:3.15}. Therefore, the calculation of the correlator proceeds exactly as in the Schwarzschild case and the fundamental element to study the presence of correlations between the Hawking quanta and their corresponding partners across the horizon, namely the $T_{uu}$ correlator, can be written in the form \eqref{eq:3.23}. The differences with the Schwarzschild spacetime are in the expressions of the null coordinates in the "in" and "out" regions ($u_{in}$ and $u$, respectively) and the relations between them. Therefore, the explicit expression of the correlator will be different for the Hayward, Simpson-Visser and Schwarzschild black holes. Nevertheless, it is possible to generalize the procedure that was used for the Schwarzschild black hole in the following way:
\begin{itemize}
    \item Compute the tortoise coordinate.
    \item Impose the continuity of the metric \eqref{eq:4.10} describing the transition from the "in" region to the "out" one on the transition null shell at $v=v_{0}$. 
    \item Obtain the relation between the retarded null coordinates outside and inside $v_{0}$, check that the function $u(u_{in})$ is invertible and compute $\frac{du}{du_{in}}$. Then, $\frac{du_{in}}{du}=\left(\frac{du}{du_{in}}\right)^{-1}$.
    \item Invert the function $u(u_{in})$. \newline
    Unfortunately, for the Hayward and Simpson-Visser black holes this can be done only in particular cases, when the relation between $u$ and $u_{in}$ can be approximated with a simpler expression. In particular, for the Hayward black hole we will consider three limits: $r^{\prime}\to r_{2}$, $r\to r_{1}$ and $r\to\infty$. For the Simpson-Visser black hole the interesting regions are $r^{\prime}\to 0$, $r\to r_{+}$ and $r\to\infty$.
    \item Write $u$ in Painlevé coordinates.
    \item Study the behavior of the density correlator in the limits mentioned above.
\end{itemize}

\subsubsection{Density correlations in the Hayward spacetime}
In the Hayward case, the calculation exploits the fact that the metric reduces to the Schwarzschild one asymptotically, while it reproduces the de Sitter metric close to the origin and to the Cauchy horizon in the $M\gg L$ limit that is considered.
When the inner point approaches the Cauchy horizon we have (see Appendix \ref{Appendix D} for details)
\begin{equation}
    \lim_{r^{\prime}\to r_{2}}\frac{1}{(1+V(r^{\prime}))^{2}}=\infty.
\end{equation}
\begin{equation}
    \lim_{r^{\prime}\to r_{2}}\frac{du^{\prime}_{in}}{du^{\prime}}=0.
\end{equation}
Therefore, one has to consider  
\begin{equation}
\label{eq:4.14}
    \lim_{r^{\prime}\to r_{2}}\frac{1}{(1+V(r^{\prime}))^{2}}\left(\frac{du^{\prime}_{in}}{du^{\prime}}\right)^{2}=4r^{2}_{2}e^{-\frac{4r_{1}}{r_{2}}+\frac{2T^{\prime}}{r_{2}}}.
\end{equation}
The fact that this expression is not vanishing on the inner horizon implies that the density correlator remains finite when one of the two points approaches this hypersurface. \newline
It is then important to verify that when $L\to 0$, which implies $r_{2}\to 0$, so that there is no Cauchy horizon but a singularity at $r=0$, the density correlator vanishes according to \eqref{eq:3.26}. \newline
Looking at equation \eqref{eq:4.14} there are two possibilities:
\begin{itemize}
    \item If $-4r_{1}+2T^{\prime}>0$, the correlator diverges when $L\to 0$.
    \item If $-4r_{1}+2T^{\prime}<0$, the correlator goes to zero when $L\to 0$.
\end{itemize}
$T^{\prime}$ is the Painlevé time it takes for the partner particle to travel the distance between the quantum atmosphere and the Cauchy horizon:
\begin{equation}
    T^{\prime}=\int_{r_{QA}}^{r_{2}}\frac{1}{V}dr=-\int_{r_{QA}}^{r_{2}}\left(\sqrt{\frac{2Mr^{2}}{r^{3}+2L^{2}M}}\right)^{-1}dr,
\end{equation}
where $r_{QA}$ is the radius of the quantum atmosphere, $r_{QA}\sim\frac{1}{k}$, with $k=\frac{3}{4M}-\frac{1}{r_{1}}$ being the surface gravity of the outer horizon of the Hayward black hole. For example, taking $M=1$ and $L=0.001$, the integral can be solved numerically, giving $T^{\prime}\simeq -4.23<0$. However, as $L$ decreases, $T^{\prime}$ increases because $r_{2}$ becomes smaller and thus it could be that $-4r_{1}+2T^{\prime}$ becomes positive when $L\to 0$. Nevertheless, this does not happen because when $L\to 0$, $T^{\prime}$ becomes the Painlevé time that it takes for the partner particle to go from the quantum atmosphere to $r=0$ in a Schwarzschild black hole, which is 
\begin{equation}
\label{eq:4.16}
    \lim_{L\to 0}T^{\prime}=\frac{8}{3}\sqrt{2}M. 
\end{equation}
Therefore,
\begin{equation}
    \lim_{L\to 0}(-4r_{1}+2T^{\prime})=\frac{-24+16\sqrt{2}}{3}M\simeq-1.37M<0.
\end{equation}
This implies that, in the $L\to 0$ limit one gets back the Schwarzschild result, as expected. \newline
In the $M\gg L$ case, the Hayward black hole differs from the Schwarzschild one only close to $r=0$. Outside the Cauchy horizon the effect of the $L$ parameter is just an infinitesimal shift of the outer horizon, which does not have any appreciable consequence. Therefore, when $r\to r_{1}$ and $r\to \infty$ every element of the density-density correlator reduces to its Schwarzschild counterpart. We can then conclude that the correlator remains finite when one of the two points hits the outer horizon and it vanishes asymptotically.

\subsubsection{Density correlations in the Simpson-Visser spacetime}
We finally discuss what happens in a Simpson-Visser non-singular black hole. In this case, the idea is to perform a Taylor expansion of the relevant quantities in the limits of interest (see Appendix \ref{Appendix E} for details). \newline
The term of the density correlator that depends only on $r^{\prime}$ remains finite when $r^{\prime}\to 0$. In fact one has
\begin{align}
\label{eq:4.18}
    &\lim_{r^{\prime}\to 0}\frac{1}{(1\pm V(r^{\prime}))^{2}}\left(\frac{du^{\prime}_{in}}{du^{\prime}}\right)^{2}=\notag\\
    &=\frac{1}{\left(1\pm\sqrt{\frac{2M}{a}}\right)^{2}}\left[\frac{\sqrt{\frac{(2m-a)^{2}}{a^{2}}T^{\prime 2}+4a^{2}}-4M}{\sqrt{\frac{(2m-a)^{2}}{a^{2}}T^{\prime 2}+4a^{2}}}\right]^{2},
\end{align}
Therefore, the density correlator remains finite and non-vanishing when the inner point approaches $r=0$ (unless the outer point reaches infinity, as it will be discussed below). In particular, $T^{\prime}$ is the Painlevé time it takes for the partner particle to go from the quantum atmosphere to $r=0$, which is given by
\begin{equation}
\label{eq:4.19}
    T^{\prime}=\int_{r_{QA}}^{0}\frac{1}{V}dr=-\int_{r_{QA}}^{0}\left(\sqrt{\frac{2M}{\sqrt{r^{2}+a^{2}}}}\right)^{-1}dr.
\end{equation}
$r_{QA}\sim \frac{1}{k}$ is the radius of the quantum atmosphere, with $k=\frac{1}{4M}\sqrt{1-\frac{a^{2}}{4M^{2}}}$ being the surface gravity of the horizon of the Simpson-Visser black hole located at $r_{+}$.
The result of the integral \eqref{eq:4.19} can be expanded for $M\gg a$,\footnote{As usual we work on a stationary geometry. Since $a$ should be a Planckian cutoff at which quantum effects of gravity become dominant, it is reasonable to assume $M\gg a$.}
\begin{equation}
    T^{\prime}\simeq\frac{8}{3}\sqrt{2}M+\frac{\sqrt{\frac{\pi}{2}}\Gamma\left(\frac{1}{4}\right)}{6\sqrt{M}\Gamma\left(\frac{3}{4}\right)}a^{3/2}+\frac{3}{4\sqrt{2}M}a^{2}.
\end{equation}
Note that, when $a=0$ one gets back the corresponding Painlevé time in the Schwarzschild black hole \eqref{eq:4.16}.
Then, one should verify that the result \eqref{eq:4.18}, with the $"+"$ sign in the first factor, reduces to the Schwarzschild one when $a\to 0$. Performing an expansion for $M\gg a$ one obtains
\begin{equation}
    \lim_{r^{\prime}\to 0}\frac{1}{(1+V(r^{\prime}))^{2}}\left(\frac{du^{\prime}_{in}}{du^{\prime}}\right)^{2}=\frac{a}{2M}+o(a^{3/2})\xrightarrow[a\to 0]{} 0.
\end{equation}
Therefore, when $a\to 0$ the density correlator vanishes and we get back the corresponding Schwarzschild result \eqref{eq:3.26}.
\newline
When one of the two points approaches the event horizon at $r=r_{+}$ one has, for $M\gg a$,
\begin{equation}
\label{eq:4.22}
    \lim_{r\to r_{+}}\frac{1}{(1+V(r))^{2}}\left(\frac{du_{in}}{du}\right)^{2}\simeq 4e^{-\frac{T}{2M}}+\frac{e^{-\frac{3T}{4M}}}{8M^{3}}\left(-12M+9Me^{\frac{T}{4M}}+2Te^{\frac{T}{4M}}\right)a^{2}+o(a^{3}).
\end{equation}
It can be noted that the zeroth order term coincides with the Schwarzschild result \eqref{eq:3.29}, as expected. In fact, far from $r=0$, the Simpson-Visser "regularization" has only the effect of shifting the horizon by an infinitesimal amount in the $M\gg a$ limit. \newline
In this spirit, when the outer point reaches infinity, the density correlator goes to zero. This can be easily checked because, in this limit, $u_{in}\to u\to \infty$ and therefore \eqref{eq:3.23} vanishes. \newline

\section{Conclusions}
\label{sec.5}
\noindent
We have studied the two-point correlation function of the density operator of a massless scalar field propagating in two non-singular black hole spacetimes to investigate the energy density correlations between the Hawking quanta escaping to infinity and their partners that enter the event horizon. The analysis focused, in particular, on the Hayward and the Simpson-Visser black holes and it was inspired by the work done for an analogue black hole generated using a Bose-Einstein condensate.  Here, a peak in the density-density correlator, which has also been measured experimentally, signals the presence of particle-partner correlations. In the Schwarzschild case this peak does not show as when a Hawking particle emerges out of the quantum atmosphere the corresponding partner has already been swallowed by the singularity and their mutual correlations are lost. We have therefore investigated the behavior of the density-density correlator when the singularity is not present considering the Hayward and Simpson-Visser non-singular black holes. This has been done generalizing the procedure already used in the Schwarzschild case. However, for these regular spacetimes it is not possible to compute analytically (mainly because the functions involved are not invertible) the correlator for any couple of points outside and inside the event horizon and so we have not been able to show the presence of a peak directly. Nevertheless, we have demonstrated that the correlator remains finite on the hypersurface that replaces the singularity, in opposition to what happens with the Schwarzschild black hole. In addition, when the black hole mass is much larger than the Planck length, the spacetime structure of the Hayward and Simpson-Visser non-singular black holes is similar to that of the Schwarzschild one apart from the region close to $r=0$ and thus one can suppose that also the form of the correlator will be qualitatively similar. However,  in this case the characteristic peak will appear due to the fact that the partner particle does not enter a singularity but it continues to travel in other regions of spacetime where the correlator is finite. Therefore, energy density correlations can propagate in other copies of our universe towards the hypersurfaces that replace the singularity in regular black hole spacetimes. To confirm these conclusions a numerical calculation of the correlator should be performed,  which could be done by generalizing the methods used in \cite{twopf1}, \cite{twopf2}. We hope to report soon on this possibility.

\appendix

\section{Appendix A}
\label{Appendix A}
\noindent
In Section \ref{sec.3} we have considered the energy density operator of a massless scalar field measured by free-falling observers in Painlevé-Gullstrand coordinates, which are obtained introducing a new time coordinate, called the Painlevé time, defined as \begin{equation}
    T=t+\int \frac{\sqrt{1-f}}{f} dr,
\end{equation}
where $f=1-\frac{2M}{r}$ for a Schwarzschild black hole.
Note that the Painlevé time coincides with the proper time of the inertial observers free-falling radially from initial zero velocity.
\newline
The two-dimensional Schwarzschild line element in these $(T, r)$ coordinates is
\begin{equation}
    ds^{2}=-fdT^{2}-2VdTdr,
\end{equation}
where $V=-\sqrt{1-f}$. The four-velocity of the observers that are considered is 
\begin{equation}
    u^{\alpha}=(1,V).
\end{equation}
($u^{\alpha}u_{\alpha}=-1$, as can be easily verified). Recalling equation \eqref{eq:3.13}, the energy density is given by 
\begin{equation}
\label{eq:A.4}
    \rho=T_{\alpha\beta}u^{\alpha}u^{\beta}=T_{TT}+2VT_{Tr}+V^{2}T_{rr}.
\end{equation}
Due to conformal invariance the trace of the stress-energy tensor is zero, $T_{\alpha\beta}=0$. Therefore, we have 
\begin{equation}
    T_{\alpha\beta}g^{\alpha\beta}=-T_{TT}-2VT_{Tr}+fT_{rr}=0.
\end{equation}
where $g^{\alpha\beta}$ is the inverse of the two-dimensional Schwarzschild metric in Painlevé-Gullstrand coordinates,
\begin{equation}
    g^{\alpha\beta}=
    \begin{pmatrix}
    -1 & -V \\
    -V & f
    \end{pmatrix}.
\end{equation}
Substituting this result in equation \eqref{eq:A.4} we obtain the following simple expression for the density operator,
\begin{equation}
    \rho=(f+V^{2})T_{rr}=T_{rr}.
\end{equation}
As discussed in Appendix \ref{Appendix B}, to calculate the two-point correlation function one has to derive twice the Wightman function \eqref{eq:3.12}. Since the latter is written in null coordinates, it is better to express also the density operator in the same coordinate system. Recalling the expression of the energy-momentum tensor of a massless scalar field \eqref{eq:3.14}, the components in null coordinates are
\begin{align}
    T_{uu}&=\partial_{u}\phi\partial_{u}\phi, \\
    T_{vv}&=\partial_{v}\phi\partial_{v}\phi, \\
    T_{uv}&=T_{vu}=0.
\end{align}
Applying the usual tensor transformation rules to $T_{rr}$ we get:
\begin{align}
\label{eq:A.11}
    T_{rr}&=\frac{\partial x^{\alpha}}{\partial r}\frac{\partial x^{\beta}}{\partial r}T_{\alpha\beta}= \notag \\
    &=\left(\frac{\partial u}{\partial r}\right)^{2}T_{uu}+\left(\frac{\partial v}{\partial r}\right)^{2}T_{vv}+2\frac{\partial u}{\partial r}\frac{\partial v}{\partial r}T_{uv}.
\end{align}
The relation between the Eddington-Finkelstein coordinates and the Painlevé-Gullstrand ones is
\begin{align}
\label{eq:A.12}
    u&=t-r^{*}=T-\int\frac{\sqrt{1-f}+1}{f}dr, \\
    v&=t+r^{*}=T-\int\frac{\sqrt{1-f}-1}{f}dr.
\end{align}
Deriving and substituting into equation \eqref{eq:A.11} one obtains the following expression for the density operator written in $(u,v)$ coordinates:
\begin{align}
\label{eq:A.14}
    \rho=T_{rr}&=\left(\frac{\sqrt{1-f}+1}{f}\right)^{2}T_{uu}+\left(\frac{\sqrt{1-f}-1}{f}\right)^{2}T_{vv}= \notag \\
    &=\frac{T_{uu}}{(1+V)^{2}}+\frac{T_{vv}}{(1-V)^{2}}.
\end{align}

\section{Appendix B}
\label{Appendix B}
\noindent
Given the expression \eqref{eq:3.18} of the density-density correlator, we want to write it in null coordinates, which in turn can be expressed in Painlevé-Gullstrand coordinates, in order to study how it behaves as a function of the position of the two points outside and inside the event horizon. 
\newline
As remarked in Section \ref{The density-density correlator}, the fundamental object to study possible energy density correlations between a Hawking particle and its partner across the event horizon of a black hole in a two-dimensional spacetime is the $T_{uu}$ correlator. Applying the Wick's theorem, it can be written as a differential operator applied to the Wightman function:
\begin{equation}
\label{eq:B.1}
 \bra{in}T_{uu}\left(x\right)T_{u^{\prime}u^{\prime}}\left(x^{\prime}\right)\ket{in}=\left(\partial_{u}\partial_{u^{\prime}}\bra{in}\phi\left(x\right)\phi\left(x^{\prime}\right)\ket{in}\right)^{2}.
\end{equation}
Then, we have to calculate
\begin{align}
\label{eq:B.2}
    \partial_{u}\partial_{u^{\prime}}\bra{in}\phi(x)\phi(x^{\prime})\ket{in}&=-\frac{1}{4\pi}\partial_{u}\partial_{u^{\prime}}\ln \frac{\left(u_{in}-u_{in}^{\prime}\right)\left(v-v^{\prime}\right)}{\left(u_{in}-v^{\prime}\right)\left(v-u_{in}^{\prime}\right)}= \notag \\ 
    &=-\frac{1}{4\pi}\frac{du_{in}}{du}\frac{du_{in}^{\prime}}{du^{\prime}}\frac{1}{\left(u_{in}-u_{in}^{\prime}\right)^{2}}, 
\end{align}
where we have used the expression \eqref{eq:3.12} for the Wightman function. Let's now recall the equation \eqref{eq:3.8}, which gives $u$ as a function of $u_{in}$. Since $x$ is taken outside the horizon, $u_{in}<0$ and we have 
\begin{equation}
\label{eq:B.3}
    u=u_{in}-4M\ln\left({-\frac{u_{in}}{4M}}\right).
\end{equation}
Since this function is invertible when $u_{in}<0$ one has $\frac{du_{in}}{du}=\left(\frac{du}{du_{in}}\right)^{-1}$.
Taking the derivative of \eqref{eq:B.3} with respect to $u_{in}$ we get
\begin{equation}
    \frac{du}{du_{in}}=\frac{u_{in}-4M}{u_{in}},
\end{equation}
from which
\begin{equation}
\label{eq:B.5}
\frac{du_{in}}{du}=\frac{u_{in}}{u_{in}-4M}.
\end{equation}
Repeating the same calculation for the "primed" sector (remember that $x^{\prime}$ is inside the horizon and thus $u_{in}^{\prime}>0$) one obtains
\begin{equation}
\label{eq:B.6}
    \frac{du_{in}^{\prime}}{du^{\prime}}=\frac{u_{in}^{\prime}}{u_{in}^{\prime}-4M}.
\end{equation}
Now we can insert the results \eqref{eq:B.5} and \eqref{eq:B.6} into the equation \eqref{eq:B.2}:
\begin{equation}
\label{eq:B.7}
    \partial_{u}\partial_{u^{\prime}}\bra{in}\phi(x)\phi(x^{\prime})\ket{in}=-\frac{1}{4\pi}\frac{u_{in}u_{in}^{\prime}}{\left(u_{in}-u_{in}^{\prime}\right)^{2}}\frac{1}{\left(u_{in}-4M\right)\left(u_{in}^{\prime}-4M\right)}.
\end{equation}
It is convenient to use the following identity
\begin{equation}
    -\frac{u_{in}u_{in}^{\prime}}{\left(u_{in}-u_{in}^{\prime}\right)^{2}}=\frac{1}{4\cosh^{2}{\ln{\sqrt{-\frac{u_{in}}{u_{in}^{\prime}}}}}}.
\end{equation}
Therefore, the $T_{uu}$ correlator \eqref{eq:B.1} finally reads
\begin{equation}
\label{eq:B.9}
    \bra{in}T_{uu}\left(x\right)T_{u^{\prime}u^{\prime}}\left(x^{\prime}\right)\ket{in}=\frac{1}{256\pi^{2}}\frac{1}{\cosh^{4}{\ln{\sqrt{-\frac{u_{in}}{u_{in}^{\prime}}}}}}\frac{1}{\left(u_{in}-4M\right)^{2}\left(u_{in}^{\prime}-4M\right)^{2}}.
\end{equation}
For completeness, we also report the calculation of the other three terms in the correlator \eqref{eq:3.18}. 
In particular, repeating the same procedure for the $v$-sector we obtain
\begin{equation}
    \bra{in}T_{vv}(x)T_{v^{\prime}v^{\prime}}(x^{\prime})\ket{in}=\left(\partial_{v}\partial_{v^{\prime}}\bra{in}\phi(x)\phi(x^{\prime})\ket{in}\right)^{2}.
\end{equation}
Using again the Wightman function \eqref{eq:3.12} we have
\begin{align}
    \partial_{v}\partial_{v^{\prime}}\bra{in}\phi(x)\phi(x^{\prime})\ket{in}&=-\frac{1}{4\pi}\partial_{u}\partial_{u^{\prime}}\ln \frac{\left(u_{in}-u_{in}^{\prime}\right)\left(v-v^{\prime}\right)}{\left(u_{in}-v^{\prime}\right)\left(v-u_{in}^{\prime}\right)}= \notag \\ 
    &=-\frac{1}{4\pi}\frac{1}{\left(v-v^{\prime}\right)^{2}}.
\end{align}
Thus, the $T_{vv}$ correlator reads
\begin{equation}
\label{eq:B.12}
    \bra{in}T_{vv}(x)T_{v^{\prime}v^{\prime}}(x^{\prime})\ket{in}=\frac{1}{16\pi^{2}}\frac{1}{\left(v-v^{\prime}\right)^{4}}.
\end{equation}
The "mixed" correlators can be computed in a similar way,
\begin{equation}
    \bra{in}T_{uu}(x)T_{v^{\prime}v^{\prime}}(x^{\prime})\ket{in}=2\left(\partial_{u}\partial_{v^{\prime}}\bra{in}\phi(x)\phi(x^{\prime})\ket{in}\right)^{2}
\end{equation}
\begin{equation}
    \bra{in}T_{vv}(x)T_{u^{\prime}u^{\prime}}(x^{\prime})\ket{in}=2\left(\partial_{v}\partial_{u^{\prime}}\bra{in}\phi(x)\phi(x^{\prime})\ket{in}\right)^{2}
\end{equation}
with 
\begin{align}
\label{eq:B.15}
    \partial_{u}\partial_{v^{\prime}}\bra{in}\phi(x)\phi(x^{\prime})\ket{in}&=
    \frac{1}{4\pi}\frac{du_{in}}{du}\frac{1}{(u_{in}-v^{\prime})^{2}}=\notag\\
    &=\frac{1}{4\pi}\frac{u_{in}}{u_{in}-4M}\frac{1}{\left(u_{in}-v^{\prime}\right)^2}
\end{align}
and
\begin{align}
\label{eq:B.16}
    \partial_{v}\partial_{u^{\prime}}\bra{in}\phi(x)\phi(x^{\prime})\ket{in}&=
    \frac{1}{4\pi}\frac{du^{\prime}_{in}}{du^{\prime}}\frac{1}{(v-u^{\prime}_{in})^{2}}=\notag\\
    &=\frac{1}{4\pi}\frac{u_{in}^{\prime}}{u_{in}^{\prime}-4M}\frac{1}{\left(u_{in}^{\prime}-v\right)^2}.
\end{align}

\section{Appendix C}
\label{Appendix C}
\noindent
In this appendix, we report the calculation of the limits of the correlator in a Schwarzschild black hole. 
\newline
Let's start considering the limit $r^{\prime}\to 0$. We have
\begin{equation}
    \lim_{r^{\prime}\to 0}V(r^{\prime})=\lim_{r^{\prime}\to 0}-\sqrt{\frac{2M}{r^{\prime}}}=\infty,
\end{equation}
and so
\begin{equation}
    \lim_{r^{\prime}\to 0}\frac{1}{(1\pm V(r^{\prime}))^{2}}=0.
\end{equation}
On the other hand, at $v=v_{0}=4M$,
\begin{equation}
    u=4M-2r^{*}
\end{equation}
Thus, using the expression \eqref{eq:3.9} for $u_{in}$,
\begin{equation}
    \lim_{r^{\prime}\to 0}u^{\prime}_{in}=-4MW(-e^{-1-\frac{|r^{*}|}{2M}})=4M,
\end{equation}
where we have used $\lim_{r^{\prime}\to 0}r^{*}=0$ and $W(-e^{-1})=-1$, considering the principal branch of the Lambert function \cite{Lambert}. This implies that \begin{equation}
    \lim_{r^{\prime}\to 0}\frac{du^{\prime}_{in}}{du^{\prime}}=\lim_{r^{\prime}\to 0}\frac{u^{\prime}_{in}}{u^{\prime}_{in}-4M}=\infty.
\end{equation}
Then, as discussed in Section \ref{sec.3}, the object that determines the behavior of the correlator in this limit is
\begin{equation}
    \lim_{r^{\prime}\to 0}\frac{1}{(1+V(r^{\prime}))^{2}}\left(\frac{du^{\prime}_{in}}{du^{\prime}}\right)^{2}.
\end{equation}
This can be recast in the following way,
\begin{equation}
    \lim_{r^{\prime}\to 0}\frac{N}{D}=\frac{0}{0},
\end{equation}
where,
\begin{equation}
    N=r^{\prime}(u^{\prime}_{in})^{2},
\end{equation}
\begin{equation}
    D=((\sqrt{r^{\prime}}-\sqrt{2M})^{2})(u^{\prime}_{in}-4M)^{2}.
\end{equation}
The idea is then to apply de l'Hopital theorem, 
\begin{equation}
    \lim_{r^{\prime}\to 0}\frac{N}{D}=\lim_{r^{\prime}\to 0}\frac{dN/dr^{\prime}}{dD/dr^{\prime}},
\end{equation}
Writing $u^{\prime}_{in}$ as a function of $u^{\prime}$ and then $u^{\prime}$ in Painlevé coordinates it is possible to compute the derivatives and then take the limit. The result is
\begin{equation}
\label{eq:C.11}
    \lim_{r^{\prime}\to 0}\frac{1}{(1+V(r^{\prime}))^{2}}\left(\frac{du^{\prime}_{in}}{du^{\prime}}\right)^{2}=0.
\end{equation}
In the $r\to 2M$ limit one has
\begin{equation}
    \lim_{r\to 2M}V(r)=\lim_{r\to 2M}-\sqrt{\frac{2M}{r}}=-1.
\end{equation}
Thus,
\begin{equation}
    \lim_{r\to 2M}\frac{1}{(1+V(r))^{2}}=\infty.
\end{equation}
On the the other hand,
\begin{equation}
    \lim_{r\to 2M}u_{in}=0,
\end{equation}
since $\lim_{r\to 2M}u=\infty$. So,
\begin{equation}
    \lim_{r\to 2M}\frac{du_{in}}{du}=\lim_{r\to 2M}\frac{u_{in}}{u_{in}-4M}=0.
\end{equation}
In this case one has to consider
\begin{equation}
\label{eq:C.16}
    \lim_{r\to 2M}\frac{1}{(1+V(r))^{2}}\left(\frac{du_{in}}{du}\right)^{2}.
\end{equation}
Looking at the expression \eqref{eq:3.20} for $u_{in}$ at late times, the derivative term can be written as
\begin{equation}
    \frac{du_{in}}{du}=\frac{e^{-\frac{u}{4M}}}{e^{-\frac{u}{4M}}+1}.
\end{equation}
From equation \eqref{eq:3.16} one obtains 
\begin{equation}
    u=T-\int\frac{\sqrt{1-f}+1}{f}dr=T-r-2\sqrt{2Mr}-4M\ln{\abs{\sqrt{\frac{r}{2M}}-1}},
\end{equation}
which can be expanded for $r\to 2M$:
\begin{equation}
\label{eq:C.19}
    u\simeq T-4M\ln{\left(\frac{r}{2M}-1\right)}.
\end{equation}
Therefore, in this limit, one has
\begin{equation}
    e^{-\frac{u}{4M}}\simeq e^{-\frac{T}{4M}}\left(\frac{r}{2M}-1\right).
\end{equation}
The limit \eqref{eq:C.16} can then be written as
\begin{equation}
    \lim_{r\to 2M}\frac{N}{D}=\frac{0}{0},
\end{equation}
where in this case
\begin{equation}
    N=r\left[e^{-\frac{T}{4M}}\left(\frac{r}{2M}-1\right)\right]^{2},
\end{equation}
\begin{equation}
    D=(\sqrt{r}-\sqrt{2M})^{2}\left[e^{-\frac{T}{4M}}\left(\frac{r}{2M}-1\right)+1\right]^{2}.
\end{equation}
Making use again of the de l'Hopital theorem, the final result is
\begin{equation}
    \lim_{r\to 2M}\frac{1}{(1+V(r))^{2}}\left(\frac{du_{in}}{du}\right)^{2}=4e^{-\frac{T}{2M}}.
\end{equation}

\section{Appendix D}
\label{Appendix D}
\begin{figure}
\centering
\includegraphics[width=11.5cm, height=5.5cm]{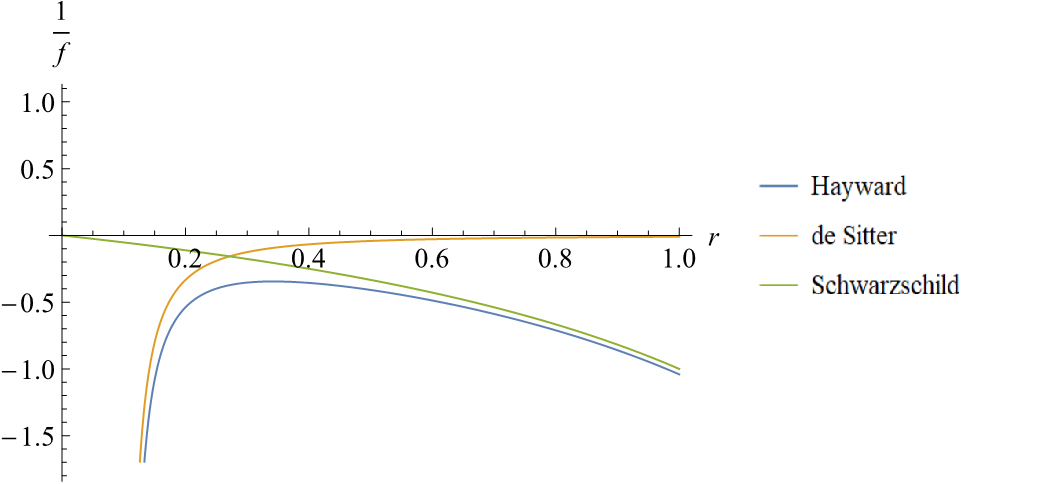}
\caption{Plot of $1/f$ for the Hayward (blue), de Sitter (orange), and Schwarzschild (green) spacetimes. In this plot $M=1$, $L=0.1$, $r_{2}=0.1025$.}
\label{fig:D.2}
\end{figure}

\noindent
We apply the procedure discussed in Section \ref{sec.4} to study the two-point correlation function of the density operator of a massless scalar field on a Hayward non-singular black hole background. In particular, we consider one point $x$ outside the outer horizon and the other $x^{\prime}$ between the two horizons. \newline
Let's start by giving the explicit expression of the tortoise coordinate,
\begin{equation}
\label{eq:D.1}
    r^{*}=\int\frac{1}{f(r)}dr,
\end{equation}
where $f(r)$ is that of equation \eqref{eq:4.4}.
Performing the integral, one obtains
\begin{align}
\label{eq:D.2}
    r^{*}&=2M\sum_{i:f(r_{i})=0}\frac{r_{i}\ln{|r-r_{i}|}}{3r_{i}-4M}+r+C\notag\\ 
    &\simeq 2M\left[\frac{r_{1}}{3r_{1}-4M}\ln{\abs{\frac{r}{r_{1}}-1}}+\frac{r_{2}}{3r_{2}-4M}\ln{\abs{\frac{r}{r_{2}}-1}}\right]+r.
\end{align}
In the last step, we have assumed $M\gg L$ in order to approximate the roots of $f(r)$ as in equations \eqref{root1} and \eqref{root2}.
We consider $r^{\prime}>r_{2}$ (which means that the inner point is taken outside of the Cauchy horizon) and so the argument of the second logarithm is always positive. \newline
It is easy to verify that for $L\to 0$, $r_{1}\to 2M$ and $r_{2}\to 0$. So, $r^{*}\to2M\ln{\abs{\frac{r}{2M}-1}}$, which is the tortoise coordinate of a Schwarzschild black hole. \newline
Next, we have to impose the continuity of the metric \eqref{eq:4.10} describing the collapse on the null shell at $v=v_{0}$, which results in the condition $r(u_{in},v_{0})=r(u,v_{0})$. We choose $v_{0}=2r_{1}$ so that the Schwarzschild results are obtained in the limit $L\to 0$. The relation between $u$ and $u_{in}$ is
\begin{equation}
\label{eq:D.3}
    u=u_{in}-4M\left[\frac{r_{1}}{3r_{1}-4M}\ln{\abs{-\frac{u_{in}}{2r_{1}}}}+\frac{r_{2}}{3r_{2}-4M}\ln{\left[\frac{2(r_{1}-r_{2})-u_{in}}{2r_{2}}\right]}\right].
\end{equation}
To compute the density correlator \eqref{eq:3.23} one needs an expression for $\frac{du_{in}}{du}$. This is easily obtained because the function $u(u_{in})$ is invertible in the subdomains $u_{in}<0$ and $u_{in}>0$ and thus $\frac{du_{in}}{du}=\left(\frac{du}{du_{in}}\right)^{-1}$. Then, one needs only to differentiate \eqref{eq:D.3} in the regions inside and outside the outer horizon, which is the surface at which $u_{in}$ changes sign. \\
For the point outside the horizon $(r>r_{1})$, $u_{in}<0$. Performing the derivative of \eqref{eq:D.3} and inverting the result, one obtains
\begin{equation}
    \frac{du_{in}}{du}=\left[1-\frac{4Mr_{1}}{3r_{1}-4M}\frac{1}{u_{in}}+\frac{4Mr_{2}}{3r_{2}-4M}\frac{1}{2(r_{1}-r_{2})-u_{in}}\right]^{-1}.
\end{equation}
For the point inside the outer horizon $(r_{2}<r^{\prime}<r_{1}$), $u^{\prime}_{in}>0$, but the derivative is the same of the point outside. \newline
Note that when $L\to 0$, $\frac{du_{in}}{du}\to \frac{u_{in}}{u_{in}-4M}$ which, once again, is the Schwarzschild result. \newline
To study the dependence of the density correlator on the position of the two points one needs, first of all, to write $u_{in}$ as a function of $u$ by inverting \eqref{eq:D.3}. Unfortunately, this function cannot be inverted analytically and thus we consider only the limits discussed in Section \ref{sec.4}. \newline
Taking into account the fact that the Hayward metric describes a particular case of a non-singular black hole with a de Sitter core, one can approximate $f(r)$ as 
\begin{equation}
    f(r)\sim\left\{
    \begin{array}{lll}
     1-\left(\frac{r}{r_{2}}\right)^{2}    & \mbox{if}    &r<r_{Q}\\
     1-\frac{r}{r_{1}}                     & \mbox{if}    &r>r_{Q}
    \end{array}
    \right.
\end{equation}
where we are still assuming $M\gg L$, so that $r_{1}\simeq 2M$ and $r_{2}\simeq L$.
Therefore, when $r\to r_{2}$, the Hayward metric can be approximated by the de Sitter one, with 
\begin{equation}
    f(r)\sim 1-\left(\frac{r}{r_{2}}\right)^{2}.
\end{equation}
Then, in this limit the tortoise coordinate can be approximated as
\begin{equation}
    r^{*}\simeq \int\frac{1}{1-\left(\frac{r}{r_{2}}\right)^{2}}dr=-\frac{r_{2}}{2}\ln{\abs{r_{2}-r}}+C.
\end{equation}
Since $r>r_{2}$ this can be written as 
\begin{equation}
    r^{*}\simeq-\frac{r_{2}}{2}\ln{\left(\frac{r}{r_{2}}-1\right)}.
\end{equation}
To make more evident the fact that this approximation is reasonable one can plot the integrand $(1/f)$ for Hayward, together with the same function for de Sitter and Schwarzschild (Fig \ref{fig:D.2}). It is evident that the de Sitter function well reproduces the Hayward one in the $r\to r_{2}$ limit.
One then has
\begin{equation}
    u=2r_{1}+r_{2}\ln{\left[\frac{2(r_{1}-r_{2})-u_{in}}{2r_{2}}\right]}.
\end{equation}
Inverting this function one obtains
\begin{equation}
\label{eq:D.10}
    u_{in}=2(r_{1}-r_{2})-2r_{2}e^{-\frac{2r_{1}}{r_{2}}+\frac{u}{r_{2}}}.
\end{equation}
It is also convenient to consider $\frac{du_{in}}{du}$ in this limit,
\begin{equation}
\label{eq:D.11}
    \frac{du_{in}}{du}=-\frac{2(r_{1}-r_{2})-u_{in}}{r_{2}}.
\end{equation}
When the outer point approaches the outer horizon $r\to r_{1}$, $f(r)$ can be approximated as
\begin{equation}
    f(r)\sim 1-\frac{r_{1}}{r},
\end{equation}
which is the same of the Schwarzschild black hole, simply with the horizon location slightly shifted from $2M$ to $r_{1}$. Therefore, in this limit, one has 
\begin{equation}
    u_{in}=-2r_{1}W(e^{-\frac{u}{2r_{1}}}).
\end{equation}
Asymptotically, when $r\to\infty$, $f(r)\to 1 $ and thus there is no distinction between $r^{*}$ and $r$. Consequently, $u_{in}=u$. \newline
Finally, in order to analyze the behavior of the density-density correlator as a function of the positions of the two points we need to write $u$ in Painlevé coordinates. This is done by approximating $f(r)$ as before in the interesting limits. \newline
In particular, when $r\to r_{2}$, one can use the de Sitter core approximation. Thus, in this limit
\begin{equation}
    \int\frac{\sqrt{1-f}+1}{f}dr\simeq \int\frac{\frac{r}{r_{2}}+1}{1-\left(\frac{r}{r_{2}}\right)^{2}}=-r_{2}\ln{(r-r_{2})}.
\end{equation}
Therefore, when $r\to r_{2}$ one has
\begin{equation}
\label{eq:D.15}
    u=T+r_{2}\ln{(r-r_{2})}.
\end{equation}
Instead, when $r\to r_{1}$ it is possible to use the Schwarzschild approximation for $f(r)$ and so $u$ is written exactly as in the Schwarzschild case, simply with $2M$ replaced by $r_{1}$. \newline
Finally, when $r\to\infty$, $f(r)\to 1 $ and thus the Painlevé time and the ordinary Schwarzschild time coincide, as well as the tortoise coordinate and the usual radial coordinate. So, in this limit,
\begin{equation}
    u=T-r.
\end{equation}
We now have all the elements to discuss the behavior of the density correlator \eqref{eq:3.23} in the $r^{\prime}\to r_{2}$ limit for the Hayward black hole. 
\begin{equation}
    \lim_{r^{\prime}\to r_{2}}V(r^{\prime})=\lim_{r^{\prime}\to r_{2}}\left(-\frac{r^{\prime}}{r_{2}}\right)=-1,
\end{equation}
and thus
\begin{equation}
    \lim_{r^{\prime}\to r_{2}}\frac{1}{(1+V(r^{\prime}))^{2}}=\infty.
\end{equation}
$\frac{du^{\prime}_{in}}{du^{\prime}}$ can be approximate as in equation \eqref{eq:D.11}, with $u_{in}$ given by \eqref{eq:D.10} and $u$ by \eqref{eq:D.15}. So,
\begin{equation}
    \lim_{r^{\prime}\to r_{2}}u^{\prime}=-\infty,
\end{equation}
which implies
\begin{equation}
    \lim_{r^{\prime}\to r_{2}}u^{\prime}_{in}=2(r_{1}-r_{2}),
\end{equation}
and 
\begin{equation}
    \lim_{r^{\prime}\to r_{2}}\frac{du^{\prime}_{in}}{du^{\prime}}=0.
\end{equation}
Then 
\begin{align}
\label{eq:D.22}
    &\lim_{r^{\prime}\to r_{2}}\frac{1}{(1+V(r^{\prime}))^{2}}\left(\frac{du^{\prime}_{in}}{du^{\prime}}\right)^{2}=\notag\\
    &=\lim_{r^{\prime}\to r_{2}}\frac{1}{\left(1-\frac{r^{\prime}}{r_{2}}\right)^{2}}\left[-\frac{2(r_{1}-r_{2})-2(r_{1}-r_{2})-2r_{2}e^{-\frac{2r_{1}}{r_{2}}}e^{\frac{T^{\prime}}{r_{2}}+\ln{(r^{\prime}-r_{2})}}}{r_{2}}\right]^{2}=\notag\\
    &=\frac{r^{2}_{2}}{(r^{\prime}-r_{2})^{2}}\left[\frac{2r_{2}e^{-\frac{2r_{1}}{r_{2}}}e^{\frac{T^{\prime}}{r_{2}}}(r^{\prime}-r_{2})}{r_{2}}\right]^{2}=\notag\\
    &=4r^{2}_{2}e^{-\frac{4r_{1}}{r_{2}}+\frac{2T^{\prime}}{r_{2}}}.
\end{align}

\section{Appendix E}
\label{Appendix E}
\noindent
According to the general procedure discussed in Section \ref{sec.4}, we start giving an explicit expression for the tortoise coordinate
\begin{align}
    r^{*}&=\int\frac{1}{f(r)}dr=\int\left(1-\frac{2M}{\sqrt{r^{2}+a^{2}}}\right)^{-1}dr=\notag\\
    &=r-\frac{8M^{2}}{\sqrt{4M^{2}-a^{2}}}\mbox{artanh}\left[\frac{2M+r-\sqrt{r^{2}+a^{2}}}{\sqrt{4M^{2}-a^{2}}}\right]+2M\mbox{artanh}\left[\frac{r}{\sqrt{r^{2}+a^{2}}}\right]+C,
\end{align}
where $C$ is an integration constant. \newline
Imposing the continuity of the metric \eqref{eq:4.10} one obtains again the condition $r(u_{in},v_{0})=r(u,v_{0})$, which, written explicitly, gives 
\begin{align}
    &\frac{v_{0}-u_{in}}{2}-\frac{8M^{2}}{\sqrt{4M^{2}-a^{2}}}\mbox{artanh}\left[\frac{2M+\frac{v_{0}-u_{in}}{2}-\sqrt{\left(\frac{v_{0}-u_{in}}{2}\right)^{2}+a^{2}}}{\sqrt{4M^{2}-a^{2}}}\right]\notag\\
    &+2M\mbox{artanh}\left[\frac{\frac{v_{0}-u_{in}}{2}}{\sqrt{\left(\frac{v_{0}-u_{in}}{2}\right)^{2}+a^{2}}}\right]+C=\frac{v_{0}-u}{2}.
\end{align}
Since $v_{0}$ is arbitrary, in this case it is convenient to choose $v_{0}=0$ to simplify the calculation. However, this is different from what we have done for the Schwarzschild and the Hayward black holes, where we have chosen $v_{0}$ to be twice the radius of the outer horizon. This affects the relation between the inner and outer retarded null coordinates, and therefore it will not be enough to take the $a\to 0$ limit to check the consistency with the Schwarzschild case in all the expressions that involve $u_{in}$ and $u$. In order to do that, we will have to write everything in Painlevé coordinates and only at that stage verify the agreement with the Schwarzschild results for $a\to 0$. \newline
We continue by writing $u$ as a function of $u_{in}$ as 
\begin{align}
\label{eq:E.3}
    u&=u_{in}+\frac{16M^{2}}{\sqrt{4M^{2}-a^{2}}}\mbox{artanh}\left[\frac{2M-\frac{u_{in}}{2}-\sqrt{\frac{u_{in}^{2}}{4}+a^{2}}}{\sqrt{4M^{2}-a^{2}}}\right]\notag\\
    &-4M\mbox{artanh}\left[\frac{-\frac{u_{in}}{2}}{\sqrt{\frac{u_{in}^{2}}{4}+a^{2}}}\right]+C.
\end{align}
The next step to study the density correlator \eqref{eq:3.23} is to compute $\frac{du_{in}}{du}$. This can be easily done since the function $u(u_{in})$ is invertible and therefore $\frac{du_{in}}{du}=\left(\frac{du}{du_{in}}\right)^{-1}$. Then, one only needs to differentiate \eqref{eq:E.3},
\begin{equation}
\label{eq:E.4}
    \frac{du_{in}}{du}=\left(\frac{du}{du_{in}}\right)^{-1}=\frac{\sqrt{u^{2}_{in}+4a^{2}}-4M}{\sqrt{u^{2}_{in}+4a^{2}}}.
\end{equation}
Remember that the aim is to discuss how the density correlator changes as we move the two points inside and outside the horizon. Therefore, we need to write $u_{in}$ as a function of the radial coordinate. The strategy is always the same: first invert \eqref{eq:E.3} and then write $u$ in Painlevé coordinates. Unfortunately, this can be done only in the limits discussed in Section \ref{sec.4}, where the relation between $u$ and $u_{in}$ can be properly approximated. \newline
Let's start considering $r^{\prime}\to 0$. \newline
In this limit, the tortoise coordinate can be expanded as
\begin{equation}
    r^{*}=C-\frac{8M^{2}}{\sqrt{4M^{2}-a^{2}}}\mbox{artanh}\left[\frac{2M-a}{\sqrt{4M^{2}-a^{2}}}\right]-\frac{a}{2M-a}r+o(r^{3}).
\end{equation}
When $a\to 0$, this should give back the Schwarzschild result, for which
\begin{equation}
    \lim_{r\to 0}r^{*}=0.
\end{equation}
However,
\begin{equation}
    \lim_{a\to 0}\left[-\frac{8M^{2}}{\sqrt{4M^{2}-a^{2}}}\mbox{artanh}\left[\frac{2M-a}{\sqrt{4M^{2}-a^{2}}}\right]\right]=\infty.
\end{equation}
So, the idea is to choose the integration constant $C$ to cancel the zeroth order term. In this way,
\begin{equation}
    r^{*}=-\frac{a}{2M-a}r+o(r^{3}).
\end{equation}
Then, in this limit, the relation between $u$ and $u_{in}$ becomes
\begin{equation}
    u=-\frac{a}{2M-a}u_{in},
\end{equation}
which gives
\begin{equation}
\label{eq:E.10}
    u_{in}=-\frac{2M-a}{a}u.
\end{equation}
When $r\to r_{+}$ one can expand $1/f(r)$ as
\begin{equation}
    \frac{1}{f(r)}\simeq\frac{4M^{2}}{\sqrt{4M^{2}-a^{2}}}\frac{1}{r-\sqrt{4M^{2}-a^{2}}}.
\end{equation}
Integrating the leading order one obtains
\begin{equation}
    r^{*}\simeq\frac{4M^{2}}{\sqrt{4M^{2}-a^{2}}}\ln{\abs{\frac{r}{\sqrt{4M^{2}-a^{2}}}-1}}.
\end{equation}
In this limit, the relation between $u$ and $u_{in}$ becomes
\begin{equation}
    \frac{4M^{2}}{\sqrt{4M^{2}-a^{2}}}\ln{\left(-\frac{u_{in}}{2\sqrt{4M^{2}-a^{2}}}-1\right)}=-\frac{u}{2},
\end{equation}
from which
\begin{equation}
    u=-\frac{8M^{2}}{\sqrt{4M^{2}-a^{2}}}\ln{\left(-\frac{u_{in}}{2\sqrt{4M^{2}-a^{2}}}-1\right)}.
\end{equation}
Inverting this expression one obtains
\begin{equation}
\label{eq:E.15}
    u_{in}=-2\sqrt{4M^{2}-a^{2}}-2\sqrt{4M^{2}-a^{2}}e^{-\frac{4M^{2}-a^{2}}{8M^{2}}u}.
\end{equation}
Finally, when $r\to\infty$, $f(r)\to 1 $ and thus $r^{*}=r$, which implies $u_{in}=u$. \newline
One has then to write $u$ in Painlevé coordinates, which can be done in the limits mentioned above.
\newline
When $r\to 0$ one has 
\begin{equation}
\label{eq:E.16}
    u=T-\int\frac{\sqrt{1-f}+1}{f}dr\simeq T+\frac{\sqrt{a}(\sqrt{a}+\sqrt{2M})}{2M-a}r.
\end{equation}
Expanding for $r\to r_{+}$ one obtains
\begin{equation}
\label{eq:E.17}
    u=T-\int\frac{\sqrt{1-f}+1}{f}dr\simeq T-\frac{8M^{2}}{\sqrt{4M^{2}-a^{2}}}\ln{\abs{\frac{r}{\sqrt{4M^{2}-a^{2}}}-1}}.
\end{equation}
When $r\to\infty$ the Painlevé time and the Schwarzschild time coincide, as well as the tortoise coordinate and the radial coordinate, and so
\begin{equation}
    u=T-r.
\end{equation}
We now have all the elements to study the density correlator in the relevant limits for the Simpson-Visser non-singular black hole. \newline
Let's start considering the limit for $r^{\prime}\to 0$. Since
\begin{equation}
\label{eq:E.19}
    \lim_{r^{\prime}\to 0}V(r^{\prime})=-\sqrt{\frac{2M}{a}},
\end{equation}
the factors $\frac{1}{(1+V(r^{\prime}))^{2}}$ and $\frac{1}{(1-V(r^{\prime}))^{2}}$ remain finite in this limit. \newline
Then, when $r^{\prime}$ is close to zero, $u^{\prime}$ can be approximated as in equation \eqref{eq:E.16}, and using the result \eqref{eq:E.10} one obtains
\begin{equation}
\label{eq:E.20}
    u^{\prime}_{in}\simeq -\frac{2M-a}{a}T^{\prime}-\frac{\sqrt{a}+\sqrt{2M}}{\sqrt{a}}r^{\prime}.
\end{equation}
This implies that the term of the density correlator that depends only on $r^{\prime}$ remains finite when $r^{\prime}\to 0$. Using the equations \eqref{eq:E.19} and \eqref{eq:E.20}, together with \eqref{eq:E.4}, one arrives at the result \eqref{eq:4.18}.
\newline
One can then study the limit for $r\to r_{+}$ of the density correlator. 
\begin{equation}
    \lim_{r\to r_{+}}V(r)=-1
\end{equation}
and thus
\begin{equation}
    \lim_{r\to r_{+}}\frac{1}{(1+V(r))^{2}}=\infty.
\end{equation}
Then, from equation \eqref{eq:E.17},
\begin{equation}
    \lim_{r\to r_{+}}u=\infty.
\end{equation}
Using equation \eqref{eq:E.15}, one has
\begin{equation}
    \lim_{r\to r_{+}}u_{in}=-2\sqrt{4M^{2}-a^{2}}
\end{equation}
and applying \eqref{eq:E.4},
\begin{equation}
    \lim_{r\to r_{+}}\left(\frac{du_{in}}{du}\right)=0.
\end{equation}
Therefore, one has to compute the following limit:
\begin{equation}
    \lim_{r\to r_{+}}\frac{1}{(1+V(r))^{2}}\left(\frac{du_{in}}{du}\right)^{2}.
\end{equation}
This can be written as 
\begin{equation}
    \lim_{r\to r_{+}}\frac{N}{D}=\frac{0}{0},
\end{equation}
where,
\begin{equation}
    N=\sqrt{r^{2}+a^{2}}\left\{\sqrt{\left[-2r_{+}-2r_{+}e^{-\frac{r_{+}}{8M^{2}}T}\left(\frac{r}{r_{+}}-1\right)\right]^{2}+4a^{2}}-4M\right\}^{2},
\end{equation}
\begin{equation}
    D=[(r^{2}+a^{2})^{1/4}-\sqrt{2M}]^{2}\left\{\sqrt{\left[-2r_{+}-2r_{+}e^{-\frac{r_{+}}{8M^{2}}T}\left(\frac{r}{r_{+}}-1\right)\right]^{2}+4a^{2}}\right\}^{2}.
\end{equation}
Applying de l'Hopital theorem and expanding for $M>>a$ one obtains the result \eqref{eq:4.22}.

\clearpage

\printbibliography

@article{Hawking:paper,
  title = {Particle creation by black holes},
  author = {Hawking, S. W.},
  journal = {Commun.Math. Phys.},
  volume = {43},
  pages = {199--220},
  year = {1975},
  doi = {https://doi.org/10.1007/BF02345020},
}

@book{Fabbri:book,
author = {Fabbri, A. and Navarro-Salas, J.},
title = {Modelling black hole evaporation},
publisher = {Imperial College Press},
date = {2005},
}

@article{Unruh,
  title = {Experimental Black-Hole Evaporation?},
  author = {Unruh, W. G.},
  journal = {Phys. Rev. Lett.},
  volume = {46},
  pages = {1351--1353},
  year = {1981},
  doi = {10.1103/PhysRevLett.46.1351},
}

@article{AcousticCorrelations,
  title = {Nonlocal density correlations as a signature of Hawking radiation from acoustic black holes},
  author = {Balbinot, R. and Fabbri, A. and Fagnocchi, S. and Recati, A. and Carusotto, I.},
  journal = {Phys. Rev. A},
  volume = {78},
  pages = {021603},
  year = {2008},
  doi = {10.1103/PhysRevA.78.021603},
}

@article{RampupB&F,
  title = {Ramp-up of Hawking Radiation in Bose-Einstein-Condensate Analog Black Holes},
  author = {Balbinot, R. and Fabbri, A.},
  journal = {Phys. Rev. Lett.},
  volume = {126},
  pages = {111301},
  year = {2021},
  doi = {10.1103/PhysRevLett.126.111301},
}

@article{Steinhauer,
  title = {Observation of quantum Hawking radiation and its entanglement in an analogue black hole},
  author = {Steinhauer, J.},
  journal = {Nature Phys},
  volume = {12},
  pages = {959--965},
  year = {2016},
  doi = {https://doi.org/10.1038/nphys3863},
}

@article{Nova,
  title = {Observation of thermal Hawking radiation and its temperature in an analogue black hole},
  author = {{de Nova}, J. R. M. and Golubkov, K. and Kolobov, V. I. and Steinhauer, J.},
  journal = {Nature},
  volume = {569},
  pages = {688--691},
  year = {2019},
  doi = {https://doi.org/10.1038/s41586-019-1241-0},
}

@article{ArticoloPrincipale,
  title = {Quantum correlations across the horizon in acoustic and gravitational black holes},
  author = {Balbinot, R. and Fabbri, A.},
  journal = {Phys. Rev. D},
  volume = {105},
  pages = {045010},
  year = {2022},
  doi = {10.1103/PhysRevD.105.045010},
}

@article{Lambert,
  title = {Mirror reflections of a black hole},
  author = {Good, M. R. R. and Anderson, P. R. and Evans, C. R.},
  journal = {Phys. Rev. D},
  volume = {94},
  issue = {6},
  pages = {065010},
  year = {2016},
  doi = {10.1103/PhysRevD.94.065010},
}

@book{Birrell&Davies:book,
author = {Birrell, N. D. and Davies, P. C. W.},
title = {Quantum fields in curved space},
publisher = {Cambridge Univiversity Press},
date = {1982},
location = {Cambridge},
}

@book{Parker&Toms:book,
author = {Parker, L. E. and Toms, D. J.},
title = {Quantum field theory in curved spacetime: quantized fields and gravity},
publisher = {Cambridge University Press},
date = {2009},
location = {Cambridge},
}

@article{GPcoordinates,
author = {Martel, K.  and Poisson, E.},
title = {Regular coordinate systems for Schwarzschild and other spherical spacetimes},
journal = {American Journal of Physics},
volume = {69},
number = {4},
pages = {476-480},
year = {2001},
doi = {10.1119/1.1336836},
}

@article{Giddings:QA,
  title = {Hawking radiation, the Stefan–Boltzmann law, and unitarization},
  author = {Giddings, S. B.},
  journal = {Phys. Rev. B},
  volume = {754},
  pages = {39--42},
  year = {2016},
  doi = {https://doi.org/10.1016/j.physletb.2015.12.076},
}

@article{Liberati:QA,
  title = {The black hole quantum atmosphere},
  author = {Dey, R. and Liberati, S. and Pranzetti, D.},
  journal = {Phys. Rev. B},
  volume = {774},
  pages = {308-316},
  year = {2017},
  doi = {https://doi.org/10.1016/j.physletb.2017.09.076},
}

@article{QA2,
title = {Black hole quantum atmosphere for freely falling observers},
author = {Dey, R. and Liberati, S. and Mirzaiyan. Z. and  Pranzetti. D.},
journal = {Physics Letters B},
volume = {797},
pages = {134828},
year = {2019},
doi = {https://doi.org/10.1016/j.physletb.2019.134828},
}

@article{Penrose,
  title = {Gravitational Collapse and Space-Time Singularities},
  author = {Penrose, R.},
  journal = {Phys. Rev. Lett.},
  volume = {14},
  pages = {57--59},
  year = {1965},
  doi = {10.1103/PhysRevLett.14.57},
}

@article{Poisson&Israel,
  title = {Structure of the black hole nucleus},
  author = {Poisson, E. and Israel, W.},
  journal = {Class. Quantum Grav.},
  volume = {5},
  number = {12},
  pages = {L201--L205},
  year = {1988},
  doi = {10.1088/0264-9381/5/12/002},
}

@article{Balbinot&Poisson,
  title = {Stability of the Schwarzschild-de Sitter model},
  author = {Balbinot, R. and Poisson, E.},
  journal = {Phys. Rev. D},
  volume = {41},
  pages = {395--402},
  year = {1990},
  doi = {10.1103/PhysRevD.41.395},
}

@article{Markov,
  title = {Possible state of matter just before the collapse stage},
  author = {Markov, M. A.},
  journal = {Pis'ma Zh. Eskp. Teor. Fiz.},
  volume = {46},
  pages = {342--345},
  year = {1987},
}

@article{Hayward:NSBH,
  title = {Formation and Evaporation of Nonsingular Black Holes},
  author = {Hayward, S. A.},
  journal = {Phys. Rev. Lett.},
  volume = {96},
  pages = {031103},
  year = {2006},
  doi = {10.1103/PhysRevLett.96.031103},
}

@article{SV:NSBH,
  title = {Black-bounce to traversable wormhole},
  author = {Simpson, A. and Visser, M.},
  journal = {JCAP},
  volume = {5},
  number = {12},
  pages = {L201--L205},
  year = {2017},
  doi = {10.1088/1475-7516/2019/02/042},
}

@article{BlackBounce,
  title = {Novel black-bounce spacetimes: Wormholes, regularity, energy conditions, and causal structure},
  author = {Lobo, F. S. N. and Rodrigues, M. E. and Silva, M. V. de S. and Simpson, A. and Visser, M.},
  journal = {Phys. Rev. D},
  volume = {103},
  pages = {084052},
  year = {2021},
  doi = {10.1103/PhysRevD.103.084052},
}

@article{Thorne,
  title = {Wormholes in spacetime and their use for interstellar travel: A tool for teaching general relativity},
  author = {Morris, M. S. and Thorne, K. S.},
  journal = {American Journal of Physics},
  volume = {56},
  number = {5},
  pages = {395--412},
  year = {1988},
  doi = {10.1119/1.15620},
}

@article{SV:Vaidya,
  title = {Vaidya spacetimes, black-bounces, and traversable wormholes},
  author = {Simpson, A. and Mart{\'{i}}n-Moruno, P. and Visser, M.},
  journal = {Classical and Quantum Gravity},
  volume = {36},
  number = {14},
  pages = {145007},
  year = {2019},
  doi = {10.1088/1361-6382/ab28a5},
}

@article{twopf1,
  title = {Two-point function of a quantum scalar field in the interior region of a Reissner-Nordstrom black hole},
  author = {Lanir, A. and Levi, A. and Ori, A. and Sela, O.},
  journal = {Phys. Rev. D},
  volume = {97},
  pages = {024033},
  year = {2018},
  doi = {10.1103/PhysRevD.97.024033},
}

@article{twopf2,
  title = {Two-point function of a quantum scalar field in the interior region of a Kerr black hole},
  author = {Zilberman, N. and Casals, M. and Ori, A. and Ottewill, A. C.},
  publisher = {arXiv},
  year = {2022},
  doi = {10.48550/ARXIV.2203.07780},
}

@article{Candelas,
  title = {Vacuum polarization in Schwarzschild spacetime},
  author = {Candelas, P.},
  journal = {Phys. Rev. D},
  volume = {21},
  issue = {8},
  pages = {2185--2202},
  year = {1980},
  doi = {10.1103/PhysRevD.21.2185},
}

@article{C&F,
  title = {Trace anomalies and the Hawking effect},
  author = {Christensen, S. M. and Fulling, S. A.},
  journal = {Phys. Rev. D},
  volume = {15},
  issue = {8},
  pages = {2088--2104},
  year = {1977},
  doi = {10.1103/PhysRevD.15.2088},
}

@article{RN,
  title = {Quantum energy momentum tensor and equal time correlations in a Reissner-Nordström black hole},
  author = {Balbinot, R. and Fabbri, A.},
  publisher = {arXiv},
  year = {2023},
  doi = {https://doi.org/10.48550/arXiv.2303.11039},
}

@article{stresstensorinRN,
  title = {Hawking radiation from extremal and nonextremal black holes},
  author = {Balbinot, R. and Fabbri, A. and Farese, S. and Parentani, R.},
  journal = {Phys. Rev. D},
  volume = {76},
  issue = {12},
  pages = {124010},
  year = {2007},
  doi = {10.1103/PhysRevD.76.124010},
}

@article{nohairtheorem,
  title = {“Golden Oldie”: Gravitational Collapse: The Role of General Relativity},
  author = {Penrose, R.},
  journal = {General Relativity and Gravitation},
  volume = {34},
  
  pages = {1141–1165},
  year = {2002},
  doi = {https://doi.org/10.1023/A:1016578408204},
}

@article{vacuumnsbh,
  title = {Vacuum nonsingular black hole},
  author = {Dymnikova, I.},
  journal = {Gen Relat Gravit},
  volume = {24},
  
  pages = {235-242},
  year = {1992},
  doi = {https://doi.org/10.1007/BF00760226},
}

@article{NSBHreview,
  title = {Spherical black holes with regular center: a review of existing models including a recent realization with Gaussian sources},
  author = {Stefano A.},
  year={2008},
  eprint={0802.0330},
  archivePrefix={arXiv},
  primaryClass={gr-qc},
  doi = {https://doi.org/10.48550/arXiv.0802.0330},
}

@article{notesnsbhs,
  title = {Notes on nonsingular models of black holes},
  author = {Frolov, Valeri P.},
  journal = {Phys. Rev. D},
  volume = {94},
  issue = {10},
  pages = {104056},
  numpages = {12},
  year = {2016},
  doi = {10.1103/PhysRevD.94.104056},
}

@article{Brout:1995rd,
    author = "Brout, R. and Massar, S. and Parentani, R. and Spindel, Ph.",
    title = "{A Primer for black hole quantum physics}",
    eprint = "0710.4345",
    archivePrefix = "arXiv",
    primaryClass = "gr-qc",
    reportNumber = "ULB-TH95-02, UMG-MG-95-01, LPTHENS",
    doi = "10.1016/0370-1573(95)00008-5",
    journal = "Phys. Rept.",
    volume = "260",
    pages = "329--454",
    year = "1995"
}

@article{Sebastiani:2022wbz,
    author = "Sebastiani, Lorenzo and Zerbini, Sergio",
    title = "{Some Remarks on Non-Singular Spherically Symmetric Space-Times}",
    eprint = "2206.03814",
    archivePrefix = "arXiv",
    primaryClass = "gr-qc",
    doi = "10.3390/astronomy1020010",
    journal = "Astronomy",
    volume = "1",
    number = "2",
    pages = "99--125",
    year = "2022"
}

\end{document}